\documentclass[11pt,a4paper,fleqn]{article}
\usepackage{amsmath}
\usepackage{amssymb}
\usepackage{amsfonts}
\usepackage{mathtools}
\usepackage[utf8]{inputenc}
\usepackage{graphicx}
\usepackage{float}
\usepackage{subfig}
\usepackage{color}
\usepackage{multirow}
\newcommand{\kri}[2]{\Gamma^{#1} {}_{#2}}

\newcommand{\p}{\partial}

\newcommand{\Rthree}{\overset{3}{\mathrm{R}}{}}
\newcommand{\Rfour}{\overset{4}{\mathrm{R}}{}}

\newcommand{\B}{{\cal B}}

\newcommand{\kolo}[1]{\vphantom{#1}\stackrel{\circ}{#1}\!\vphantom{#1}}
\newcommand{\dtwo}{\kolo{\Delta}}

\newcommand{\comment}[1]{}

\newcommand{\dd}[1][]{\mathrm{d}{#1}}

\newcommand{\Rie}[2]{\Rfour^{#1} {}_{#2}}

\newcommand{\x}{{\bf x}}
\newcommand{\X}{{\bf X}}
\newcommand{\y}{{\bf y}}
\newcommand{\Y}{{\bf Y}}

\newcommand{\g}{{\rm g}}
\newcommand{\T}{{\rm T}}
\newcommand{\N}{{\rm N}}
\renewcommand{\P}{\rm P}

\newcommand{\alteration}{}

\newcommand{\ctg}{\mathop {\rm ctg}\nolimits }
\DeclareMathOperator{\mon}{mon}
\DeclareMathOperator{\dip}{dip}

\allowdisplaybreaks

\title{Gauge-invariant description of weak gravitational field on a spherically symmetric background with cosmological constant}
\author{Piotr Waluk\thanks{E-mail: Piotr.Waluk@fuw.edu.pl}\\
Center for Theoretical Physics,\\
Polish Academy of Sciences, \\
Al. Lotników 32/46, 02-668 Warsaw, Poland\\ \ \\
Jacek Jezierski\thanks{E-mail: Jacek.Jezierski@fuw.edu.pl}\\
Department of Mathematical Methods in Physics, \\
Faculty of Physics, University of Warsaw,\\
ul. Pasteura 5, 02-093 Warsaw, Poland}

\begin{document}
\numberwithin{equation}{section}

\maketitle

%\tableofcontents

\begin{abstract}
%Linear gravitational field is analyzed in a quasi-local way.
We present a formalism for analysis of linear Cauchy data on a Kottler metric. {\alteration Our approach is based on the ADM formulation of the problem of evolution.
It removes redundancy due to gauge transformations and geometric constraints. A set of four gauge-invariant, scalar functions on the Cauchy surface is produced and shown to contain full physical information from the initial data. The symplectic form of the theory and equations of motion are reformulated in terms of these invariants and an expression for the energy and angular momentum of the perturbation is produced. We also obtain a basic classification of stationary solutions.}
\end{abstract}

\section{Introduction}

Linearized Einstein equations have been investigated over the years in hopes to gain some understanding in situations where full nonlinear equations prove too difficult to solve, such as the stability problem of Schwarzschild (beginning with papers such as those of Regge and Wheeler \cite{reggewheeler}) or attempts at localization of gravitational energy \cite{lingrav}. Another important application is the description of gravitational waves, a particularly hot topic due recent success in experimental detection. As gravitational radiation is usually very weak, a perturbative approach proves appropriate.

{\alteration

Linearized gravity possesses a well-known gauge freedom of infinitesimal coordinate changes. Though sometimes helpful in simplifying the calculations, it poses an obvious problem in determining the physical relevance of the results.
Various approaches to this issue have been taken in literature, the most common being perhaps simply choosing and fixing a particular gauge.
A very popular gauge choice for the spherically symmetric case is the original choice of Regge and Wheeler \cite{reggewheeler}. It was later successfully applied to the case with cosmological constant \cite{GN} and remains useful when passing to higher order perturbations  \cite{rostworowski}, \cite{BMT}.
A problem with this approach, apart from the obvious complication of dealing with additional constraints, is that different choices of the gauge may be necessary for different parts of the calculation. A particular choice may obscure some important characteristic of the data, such as its asymptotic behavior. A situation of this kind is encountered e.g. in \cite{rostworowski}.

An alternative method is to seek gauge-invariant quantities --- expressions built of initial data components which transform trivially with respect to the gauge. These can be found through inspection of the structure of the gauge transformations. Through this direct approach one can obtain a useful formalism for analyzing dynamical systems \cite{praca}, \cite{COS} and it turns out to produce surprisingly general results in the class of maximally symmetric spacetimes in any dimension \cite{kodama}.
Another way to obtain gauge-invariant quantities is to linearize appropriate expressions vanishing on the background \cite{SW}. This trick has been successfully implemented in a proof of stability of Schwarzschild-de Sitter spacetimes \cite{nonmodal1}, \cite{nonmodal2}, to avoid decomposing the perturbation into spherical harmonics.

The above methods are often combined together in particular applications, such as a recent proof of linear stability of the Schwarzschild metric \cite{DHR}, utilizing the Newman--Penrose tetrad formalism (double null coordinates) and spinor methods. This formalism also proved effective in general considerations of existence of gauge-invariant descriptions of perturbations on spacetimes \cite{SW}.

The aim of this paper is to gain some insight into the structure and physical properties of the linear perturbations of the Kottler metric. We consider the linearized Cauchy problem for the Einstein equation in the Arnovitt--Deser--Misner (ADM) canonical formulation \cite{ADM} and propose a way to reduce the initial data to a set of four gauge-invariant scalar functions. This article represents therefore the second of the aforementioned approaches.
The distinguishing feature of our invariants is their close connection to the dynamical structure of the ADM theory through its symplectic form. In this way they constitute a ``natural'' choice and produce elegant formulae for the Hamiltonian and angular momentum density of the perturbation. Although similar invariant scalars have been derived in several papers \cite{COS}, \cite{nonmodal1}, their relation to the symplectic structure of gravity seems to remain unnoticed.

We wish to stress that no choice of gauge is made during our construction --- the invariants can be calculated directly from components of the perturbation.
They also take into account the existence of Gauss--Codazzi constraints and therefore are in one-to-one correspondence with the true (unconstrained) physical degrees of freedom of the theory.
Although our derivation is based on the $2+1$ splitting of the Cauchy surface, with respect to the spherical symmetry of the background, it does not require the splitting of data into spherical harmonics --- the whole theory is therefore contained within a simple and elegant set of four functions and four equations of motion, although an attempt at solving these equations may still require one to perform the splitting.

This scheme of describing linear perturbations was first developed in \cite{praca} for a purely Schwarzschild background. We have discovered that a significant part of these results can be generalized to a situation when a non-vanishing cosmological constant is present, with hardly any modifications. This observation was briefly communicated in the proceedings of the 3rd Conference of the Polish Society on Relativity \cite{potor}. In this paper we expand on the topic, providing a deeper commentary on the derivation of our invariant quantities and the role of their mono-dipole part. We also thoroughly discuss their relation to the symplectic structure of the theory and the resultant expression for energy. As a simple application of our framework, we make an attempt to classify stationary solutions of the evolution equations.
}

This paper is structured as follows: the first section contains a brief review of the linearized Cauchy problem and the Kottler metric. It also introduces our conventions and notation.

Section 2 contains the main results of this paper: derivation and analysis of gauge-invariant quantities, reformulation of the Cauchy problem in terms of these quantities, a study of their relation with the symplectic form of the ADM formulation {\alteration and the derivation of expressions for energy and angular momentum of the dynamic part of the perturbation.}

Section 3 presents the discussion of stationary solutions.

Section 4 explains the way to recover the original initial data from their gauge-invariant description. Explicit formulae are given.

Appendix \ref{splitting} lists various important equations of the theory in the $2+1$ splitting with respect to the geometry of the natural spherical foliation of the background.

Appendix \ref{metryka} contains the values of connection coefficients and curvature tensors for our background metric.

\subsection{Topology and notation}

Our goal is to analyze small perturbations of a spherically symmetric four-dimensional space-time. We will do it within the framework of the Cauchy problem for the Einstein equation. The following topology is therefore a natural choice: let our space-time be foliated by hypersurfaces labeled by the time coordinate (the ``Cauchy surfaces'') and each one of them be a union of spheres:
\begin{equation}
\Sigma_s=\{ x^0 = s, r_0 \leq x^3 \leq r_\infty \} = \underset{r \in [r_0,r_\infty [}{\bigcup} S_s(r),
\qquad
S_s(r)= \{ x \in \Sigma_s : x^3=r  \}.
\label{foliacja2}
\end{equation}
We are assuming our perturbations to be small enough for this topological picture to remain valid. The span of $r$ should fulfill $0 \leq r_0 < r_\infty \leq \infty$, but may be freely specified apart from that. By choosing a pair of angular coordinates on the spheres $S_s(r)$ we obtain a full coordinate system, which we arrange in the following way: $(x^0,x^1,x^2,x^3) = (\,t\,,\,\vartheta\,,\,\varphi\,,\,r\,)$. We will be working within several nested geometry levels. It is therefore useful to distinguish them by a following indexing convention: Greek letters ($\alpha, \beta, \gamma, ...$) correspond to a full set of coordinates and the four-dimensional geometry of the whole spacetime, with $;$ denoting the covariant derivative. Small Latin indices ($a, b, c, ...$) denote objects from the three-dimensional geometry of the Cauchy hypersurfaces, with appropriate covariant derivative denoted by~$|$. Big Latin letters ($A,B,C, ...$) and a covariant derivative symbol $||$ correspond to internal geometry on the surfaces of spheres $S_s(r)$. For curvature tensors the dimension of geometry they correspond to is additionally marked with a number over the tensor symbol, as it is not always obvious from the indices.

\subsection{ADM formulation of the Cauchy problem}
\label{daneADM}
Having divided the spacetime into hypersurfaces, one can interpret the Einstein equation as an equation for evolution of certain geometric data between neighboring slices. Out of various ways of formulating this problem, we choose the ADM approach, which is based on the Hilbert--Palatini variational principle. In this approach initial data on the Cauchy surface consists of a three-dimensional metric tensor and the so-called ADM momentum --- symmetric tensor density canonically conjugate to the metric:
\begin{equation*}
(  \g_{kl}, \mathrm{P}^{kl} ), \quad
\g_{kl}=\g_{\mu\nu}|_{\Sigma_s},
\quad \mathrm{P}^{kl}=\sqrt{  \g}( \g^{kl} K - K^{kl} ),
\quad \g:=\det  \g_{kl}.
\end{equation*}
In the above formula $K^{kl}$ and $K$ denote the extrinsic curvature of the Cauchy surface and its three-dimensional trace.

The Einstein equation can be now split into four Gauss--Codazzi constraints:
\begin{equation}
\label{ww}
\P_i{^l}{_{| l}} = 8\pi\sqrt{\g} \, \T_{i\mu} n^{\mu},
\end{equation}
\begin{equation}
\label{ws}
 \g \Rthree- \P^{kl}\P_{kl}
 + \frac{1}{2} \P^2  =
 16\pi  \g \T_{\mu \nu}n^{\mu}n^{\nu} + 2\Lambda  \g,
\end{equation}
and twelve equations of motion for the components of initial data:
\begin{equation}
\label{gdot}
\dot\g_{kl}=\frac{2\N}{\sqrt\g}\left( \P_{kl} -\frac 12
\g_{kl} \P \right) +
\N_{k|l} +\N_{l|k},
\end{equation}
\begin{equation}
\begin{aligned}
\dot\P_{kl}=
& -\N\sqrt\g {\Rthree}_{kl} + \sqrt\g \left( \N_{|kl} -\g_{kl}\N^{|m}{_{|m}} \right)  - \frac{2\N}{\sqrt\g}\left( \P_k{}^{m}\P_{ml}- \frac{1}{2} \P \P_{kl} \right) \\[-0.1cm]
\hspace*{-2cm}
& + \frac{\N}{2\sqrt\g}\g_{kl} \left( \P^{mn}\P_{mn} - \frac{1}{2}\P^2 \right) + \frac 12 \N\sqrt\g \g_{kl}{\Rthree} + \left( \P_{kl} \N^m \right)_{|m}  \\
\hspace*{-2cm}
&-\N_k{_{|m}} \P^m{}_l-\N_l{_{|m}} \P^m{}_k  + 8\pi {\rm N}\sqrt{\rm g}\, {\rm T}_{kl} - {\rm N} \sqrt{{\rm g}}\Lambda {\rm g}_{kl}.
\end{aligned}
\label{Pdot}
\end{equation}
Here $n^\mu$ is the normal vector of the Cauchy surface and $N:=\frac{1}{\sqrt{-\g^{00}}}$ and $N^k:=\g_{0k}$, called the ``lapse'' and ``shift'', are parameters corresponding to the freedom in gluing together consecutive hypersurfaces, see e.g. \cite{gravitation}.

%It remains a separate question what part of the whole four-dimensional geometry of the space-time can actually be recreated in this way from a single slice. As the goal of this paper is just to establish a tool-set for analysis, we do not address this topic further.

\subsection{The Kottler Metric}
\label{Kottler}

As the background for perturbations, we will be using the Kottler metric:
\begin{equation}
\eta=-f\dd[t]^2 + \frac{1}{f} \dd[r]^2+ r^2\left[\dd[\vartheta]^2+\sin^2\vartheta\dd[\varphi]^2 \right], \qquad
f(r)=1-\frac{2m}{r}-\frac{r^2}{3}\Lambda.
\label{tlo}
\end{equation}
It is a spherically symmetric vacuum solution of the Einstein equation with cosmological constant, for which we use the following convention:
\begin{equation}
\label{einstein}
2\Rfour_{\mu\nu}- \Rfour \g _{\mu\nu} + 2\Lambda   \g_{\mu\nu}= 16\pi \T_{\mu \nu}.
\end{equation}

The Kottler metric is a ``general solution'' in the sense that all spherically symmetric vacuum solutions of (\ref{einstein}) are (at least locally) isometric to either a region of (\ref{tlo}) or a region of the Nariai metric. This result is known as the generalized Birkhoff's theorem \cite{birkhoff}.

Minkowski, Schwarzschild and (Anti)de Sitter metrics are all contained in the Kottler metric as special cases, when one or both of the parameters $m$ and $\Lambda$ vanish. The Kottler metric can be in general thought of as Schwarzschild black hole located in a space-time curved by a presence of a cosmological constant. Exact properties of this solution depend on the sign of $\Lambda$ and its relation to the mass parameter $m$.

For positive (repelling) $\Lambda$, fulfilling $0<\Lambda<(3m)^{-2}$, two horizons exist in the space-time --- an event horizon of the central black hole at radius $r_S$ and a cosmological horizon at radius $r_C$, analogous to the one present in the de Sitter metric. The exact values of $r_S$ and $r_C$ depend on the values of $m$ and $\Lambda$, but the following inequalities are always fulfilled: $2m<r_S<3m$, $r_C>3m$. Due to this, a Kottler metric with such parameters is often called a Schwarzschild--de Sitter metric.

As $\Lambda$ grows in a Schwarzschild--de Sitter metric, the values of $r_S$ and $r_C$ approach each other. They coincide for $\Lambda=(3m)^{-2}$. However, the spatial distance between horizons does not tend to $0$ in this case. An appropriate rescaling of the radial coordinate while sending $\Lambda$ to $(3m)^{-2}$ allows one to obtain a metric known as the Nariai solution.

For a negative (attracting) $\Lambda$ only one horizon exists --- the black hole horizon at some $r_S$ between $0$ and $2m$. The situation resembles the Schwarzschild metric in this way. In analogy to the case above, such metric is often called the Schwarzschild--Anti de Sitter metric.

A wider description of these metrics can be found in: \cite{Kottler}, \cite{Kottler2}, \cite{notatki} and \cite{nariai}. In appendix~\ref{metryka} we listed formulae for the Levi-Civita connection and curvature tensors of the Kottler metric, for the readers convenience.

Because we are using the Cauchy problem approach to the Einstein equation, we should note that the Kottler metric, or the part of it that is contained between $r_0$ and $r_\infty$, may not be globally hyperbolic, depending on the values of $m$ and $\Lambda$. We will not concern ourselves with this problem, however, as it will not be important in most of our considerations.

\subsection{Linearized gravity}
\label{gravlin}

From this point on we restrict ourselves to the vacuum case (${\rm T}_{\mu\nu}=0$). We assume that we are in possession of some solution $\g$ to the Einstein equation (\ref{einstein}), which is a small perturbation of the Kottler metric (\ref{tlo}).
The difference of ADM data is easy to calculate, because, for our choice of coordinates, the background yields an ADM data set of very simple form: $(\eta_{kl},0)$.
The perturbation of ADM momentum is therefore equal in value to the ADM momentum of $\g$:
\begin{equation}
h_{kl}:= {\rm g}_{kl} - \eta_{kl}, \qquad P^{kl}={\rm P}^{kl}.
\label{rozklad2}
\end{equation}

We now perform a standard linearization procedure, expanding equations from section \ref{daneADM} in terms of ADM data perturbation and restricting them to first order terms. We remind the reader that all raising and lowering of indices and all covariant derivatives from this point on are calculated with respect to the background metric and dimensional restrictions thereof.
One therefore should be wary of index positions at the time of linearization. Some notes on this matter can be found in \cite{peeling}. As ``natural'' index positions for the ADM data we take those given in (\ref{rozklad2}). All further equations will be expressed in terms of these tensors, to avoid ambiguities.

Linearization of (\ref{einstein}) gives:
\begin{equation}
h_{\mu \alpha}{_{;\nu}}{^{;\alpha}} +
h_{\nu \alpha}{_{;\mu}}{^{;\alpha}} -
h_{\mu \nu}{^{;\alpha}}{_{\alpha}} -
h_\alpha{}^\alpha{_{;\mu \nu}} - \eta _{\mu \nu}
[ h_{ \alpha\beta}{^{;\alpha \beta}} -
h_{\alpha}{^{\alpha ; \beta}}{_{\beta}} ]
+ \Lambda h_\alpha{}^\alpha \eta_{\mu\nu} -
 2\Lambda h_{\mu\nu}
= 0.
\label{einstein_lin}
\end{equation}
The assumption ${\rm T}_{\mu\nu}=0$ is already taken into account.
The Gauss--Codazzi constraints in their linear form:
\begin{equation} \label{wwl}
  P_l{^k}{_{| k}}  =0,
\end{equation}
\begin{equation}   \label{wsl}
  \left( h^{kl}{_{|l}}- h^{|k} \right)_{|k} -h^{kl}\Rthree_{kl}  =0.
\end{equation}
$\Rthree_{kl}$ denotes here the three-dimensional Ricci tensor of the restriction of the background metric (\ref{tlo}) to $\Sigma_s$ and $h$ --- the trace of $h_{kl}$ with respect to it, $h:=h_{kl}\eta^{kl}$.

Let us note that the correction containing explicit $\Lambda$ in the scalar constraint (\ref{ws}) vanished in the approximation, leaving only implicit dependence on the cosmological constant through curvature and covariant derivatives. The form of the equation is indistinguishable from the pure Schwarzschild case \cite{praca}.

Passing to the linearization of the equations of motion, let us first introduce some additional notation. Let $N:=(-\eta^{00})^{-1/2}=\sqrt{f}$ be the lapse of the background. The shift of the background is identically zero. By $n:=\frac{1}{2} \sqrt{f}h^0{}_0$ we denote the perturbation of the lapse function and by $\eta:=\det \eta_{kl}$ the density defined by $\eta|_{\Sigma_s}$.
From (\ref{gdot}) and (\ref{Pdot}) we now get:
\begin{equation}
\dot{h}_{kl}  = \frac{2N}{\sqrt\eta} \left( P_{kl} -\frac 12
\eta_{kl} P \right) + h_{0k|l} +h_{0l|k} \, ,
\label{hdot}
\end{equation}
\begin{equation}
\begin{aligned}
\frac 1{\sqrt\eta}\dot{P}_{kl} =& -n \Rthree_{kl} -N\delta R{_{kl}}
  + n_{|kl} -N_{|m}\delta\Gamma^m{_{kl}} + N\Lambda h_{kl}
\\  &
   - \eta_{kl}\left( n^{|m}{_{|m}}
-\eta^{ij}\delta\Gamma^m{_{ij}}N_{|m} -h^{mn}N_{|mn}  \right),
\label{Pdotlin}
\end{aligned}
\end{equation}
where some terms were grouped for clarity:
\begin{equation*}
\delta\Gamma^m{_{kl}} :=\frac12\left(
h^m{_{k|l}}+h^m{_{l|k}}-h_{kl}{^{|m}} \right)\, ,
\end{equation*}
\begin{equation*}
\delta R{_{kl}} :=\frac12\left(
h^m{_{k|lm}}+h^m{_{l|km}}-h_{kl|m}{^{m}} -h_{|kl} \right)\, .
\end{equation*}

\subsection{Gauge transformations}
\label{trans}

Linearized gravity possesses a well-known gauge freedom of the ``infinitesimal coordinate change'', acting on the perturbation of the full metric:
\begin{equation}
h_{\mu \nu} \rightarrow h_{\mu \nu} + \xi _{\mu ; \nu} + \xi _{\nu ; \mu} \quad ( =  h_{\mu \nu} + \pounds_{\xi}\eta_{\mu\nu}) \, .
\label{gauge}
\end{equation}

The linearized Einstein equation is of course invariant with respect to this transformation, and so are the derived constraint and motion equations (\ref{wwl})---(\ref{Pdotlin}). Using (\ref{hdot}), one can easily find the way in which the gauge acts on the perturbed ADM momentum.
It turns out that for our choice of the background (diagonal and static), the parametrization of the gauge field $\xi_\mu$ splits into a three-dimensional field $\xi^k$, tangent to $\Sigma_s$, which acts on the perturbed restricted metric, and a function $\xi^0$, which governs the transformation of ADM momentum:
\begin{align}
 h_{kl} \quad \rightarrow & \quad h_{kl} + \xi_{l | k}+ \xi_{k | l},
\label{hxi}
\\
2f\Pi^{-1} P_{kl} \quad  \rightarrow & \quad 2f\Pi ^{-1} P_{kl}
 +(f \xi^0_{|k})_{|l}+(f \xi^0_{|l})_{|k} - 2 \eta_{kl} (f\xi{^{0|m}})_{|m}.
\label{pxi}
\end{align}
Here $\Pi:= \sqrt{f}\sqrt{\det \eta_{kl}}=r^2\sin \vartheta$.

\section{Construction and properties of {\alteration unconstrained}, invariant degrees of freedom}
\label{niezm}

To be able to effectively analyze properties of the weak gravitational field, it would be useful to separate true degrees of freedom from the gauge-dependent ones, and those restricted by the Gauss--Codazzi equations.

\subsection{2+1 splitting of the data, degrees of freedom of the perturbation}
\label{DataSplitting}
To construct our set of gauge-invariant degrees of freedom, we will decompose the perturbation data into scalar functions using the geometric structure of the two-spheres $S_s(r)$ (\ref{foliacja2}), which foliate our Cauchy hypersurface. {\alteration We begin by extracting two-dimensional traces: }
\begin{align*}
H:=\eta^{AB}h_{AB}, & \qquad \chi_{AB}:=h_{AB}-\frac{1}{2}\eta_{AB}H, \\
S:=\eta^{AB}P_{AB}, & \qquad  S_{AB}:= P_{AB} -\frac 12 \eta_{AB} S.
\end{align*}

{\alteration
We can now separate the metric and momentum tensors into scalar, vector and tensor parts, and further split them into scalar quantities by applying rotation and divergence operators. The resulting functions are listed in the table below:\\[1ex]
}

\begin{tabular}{ l | c | c |}
\cline{2-3}
& Even & Odd \\
&(Polar)&(Axial)\\
\hline
\multicolumn{1}{|l|} {\multirow{2}{*}{ Scalar } }& \multirow{2}{*}{$h^3{}_3\: , H \: , P^3{}_3 \: , S$} & \multirow{2}{*}{-} \\
\multicolumn{1}{|l|} {}&  &  \\
\hline
\multicolumn{1}{|l|}{Vector}& \multirow{2}{*}{$ h_3{}^A{}_{||A} \: , \; P^3{}^A{}_{||A} $} & \multirow{2}{*}{$h_{3A||B}\varepsilon^{AB}, \; P^{3A||B}\varepsilon_{AB}$ } \\
\multicolumn{1}{|l|}{$h^3{}_A, \; P^3{}_A$} &  & \\
\hline
\multicolumn{1}{|l|}{Tensor}& \multirow{2}{*}{ $\chi^{AB}{}_{||AB} \: , \; S^{AB}{}_{||AB}$ }
 & \multirow{2}{*}{$\chi^{C}{}_A{}_{||CB}\varepsilon^{AB} \: , \; S^{C}{}_A{}_{||CB}\varepsilon^{AB}$}\\
\multicolumn{1}{|l|}{$\chi_{AB}, \; S_{AB}$}  &   &   \\
\hline
\end{tabular}\\ \\

This decomposition preserves all information encompassed in the pair $(h_{kl}, P^{kl})$. Proof of this fact can be found in \cite{COS}.

This table already contains strong suggestions concerning the form of invariants that we should seek. Our twelve functions are subject to four constraint equations and are acted upon by a four-parameter group of gauge transformations. We therefore expect to obtain four physical degrees of freedom. Furthermore, we can make use of the symplectic structure of the theory and pair the degrees of freedom by means of equations of motion (which are gauge-invariant!). It is also a natural choice in this context to separate metric perturbations from momentum perturbations. Finally, note that only eight of the functions in the table are true scalar functions (``even'' or ``polar'' degrees of freedom). The other four are pseudo-scalar (``odd'', ``axial''), i.e.~dependent on our choice of manifold orientation. Degrees of freedom of different parity decouple in linear theory, further dividing our set of functions and narrowing down our reasonable choices for invariant combinations.

{\alteration
To summarize --- if we split the initial data into the odd and even parts and separate metric and momentum perturbations (which is roughly equivalent to separating the spatial parts of the 4-metric from the rest) we obtain four sets of degrees of freedom and expect each one of those sets to reduce to a single invariant quantity. These, rather natural, assumptions produce a rigid restriction on the form of invariant quantities.
}

Constraints, equations of motion and gauge transformations can also be split in this scheme. We list the resultant formulae in appendix \ref{splitting} and refer to this list when necessary, as some of the expressions are rather long.

\subsection{Axial dynamics}

We begin by discussing the axial degrees of freedom. It is the simpler of the two cases and construction of invariants is pretty straightforward. Let us begin with ADM momentum components that belong to this regime: $P^{3A||B}\varepsilon_{AB}$ and $S^{C}{}_A{}_{||CB}\varepsilon^{AB}$. These two functions are not independent, as can be seen by acting with a rotation operator on appropriate part of the vector constraint (\ref{wwA}):
\begin{equation}
\label{rotww}
(r^2 P^{3A||B}\varepsilon_{AB})_{,3}+ r^2  S^{C}{}_A{}_{||CB}\varepsilon^{AB} = 0.
\end{equation}
The component $S^{C}{}_A{}_{||CB}\varepsilon^{AB}$ is therefore entirely redundant. Furthermore, from equation (\ref{pxi3A}) we see that $P^{3A||B}\varepsilon_{AB}$ is already an invariant quantity!

To obtain its conjugate counterpart we simply calculate the time derivative. Applying the rotation operator to (\ref{P3C}) gives us:
\begin{equation}
\dot{P}^{3A||B}\varepsilon_{AB} = \frac{f}{2r^2}\left[  \Pi(\dtwo +2)h_{3A||B}\varepsilon^{AB}
 -\Pi (r^2 \chi^C{_{A||CB}}\varepsilon^{AB}),{_3} \label{protdot}\right],
\end{equation}
where $\dtwo$ is the two-dimensional Laplace operator on a unit sphere. Alternatively, we could just look for an invariant combination of $h_{3A||B}\varepsilon^{AB}$ and $\chi^{C}{}_A{}_{||CB}\varepsilon^{AB}$. As their gauge transformations are defined by a single function, $\xi_{A||B}\varepsilon^{AB}$, a short search would once again lead us to some function of the expression contained within square brackets above.

We therefore propose the following set of axial invariants, formally identical to those defined in \cite{praca}:
\begin{align}
 \y&:=2\Pi^{-1}r^2 P^{3A||B}\varepsilon_{AB}, \label{y}\\
 \Y&:= \Pi(\dtwo +2)h_{3A||B}\varepsilon^{AB}
 -\Pi (r^2 \chi^C{_{A||CB}}\varepsilon^{AB}),{_3} .\label{Y}
\end{align}
We introduced density coefficients here to switch the roles of ``positions'' and ``momenta''. This change will be justified later, upon closer inspection of information carried by these quantities.

$\Y$ and $\y$ satisfy the following set of reduced equations of motion:
\begin{align}
\label{py} \dot\y&=\frac{f}\Pi \Y , \\
\label{pY} \dot\Y&= \Pi \left\{\partial_3 \left[\frac f {r^2}(r^2\y),{_3}\right] +\frac1{r^2}(\dtwo +2) \y\right\}.
\end{align}
Equation (\ref{py}) is, of course, just equation (\ref{protdot}), rewritten in new variables. Equation (\ref{pY}) is obtained from (\ref{h3A}), (\ref{hAB}) and the vector constraint (\ref{wwA}).

We may combine these equations into a deformed wave equation for $\y$:
\begin{equation}
\left( \square +\frac{8m}{r^3}- \frac{2}{3}\Lambda\right) \y=0. \label{rfy}
\end{equation}
The symbol $\square$ denotes the d'Alembert operator, calculated with respect to the four-dimensional background $\eta_{\mu\nu}$.
We may further rewrite (\ref{rfy}) into a form resembling the famous Regge--Wheeler equation:
\begin{equation}\label{RWE}
 -\ddot\y +\frac fr\left[f(r\y)_{,3}\right]_{,3} =  V^{(-)}  \y ,\qquad V^{(-)}= -\frac f{r^2} \left(\dtwo+\frac{6m}r\right).
\end{equation}

Let us decompose $\y$ into spherical harmonics: $\y = \exp (i \sigma t)Y_l(\theta,\phi) Z^{(-)}(r)/r$ and introduce alternative radial coordinate $r^*$, defined as a solution to:  $\frac{d r^*}{d r}=1/f$. This allows us to compact the above formula into an elegant form:
\begin{equation}
\label{RWEsplit}
 \left(\frac{d^2}{dr^*{}^2}+\sigma^2 \right)Z^{(-)}Y_l=V^{(-)}Z^{(-)}Y_l.
\end{equation}
It is now easy to make a quick comparison of some properties of this equation for positive, negative and vanishing $\Lambda$. The potential $V^{(-)}$ always vanishes on the Schwarzschild horizon. However, its behavior near infinity depends on the cosmological constant, as $ \lim _{r\to \infty} V^{(-)}=-\frac 13 \Lambda l(l+1)$. It is also worth noting that in the case of positive $\Lambda$, when a cosmological horizon appears, the potential vanishes for $r=r_C$.

\subsection{Polar dynamics}
\label{dynpol}

The polar part of the data presents a significantly more complex problem in the search for invariant description. A greater number of metric and momentum components and a larger gauge group obscure the picture, and it turns out that the intended construction of a pair of functions, locally dependent on the ADM data components and fully describing the perturbation, is actually impossible.

However, a certain compromise is available. In \cite{praca}, for a Schwarzschild background, the following pair of invariant quantities was proposed:
\begin{align}
 \x& := r^2\chi^{AB}{_{||AB}}-\frac{1}{2}(\dtwo+2) H+
   \B \left[ 2 h^{33}+2r h^{3C}{_{||C}} -rf H,{_3}  \right],
                \label{x}\\
 \X &:=
 2r^2 S^{AB}{_{||AB}} + \B \left[ 2rP^{3A}{_{||A}}
+\dtwo P^3{_3} \right] , \label{X}
\end{align}
which also turn out to be gauge-invariant in case of a Kottler background.
The symbol $ \B $ denotes the following operator:
\begin{equation}
 \B := (\dtwo +2)\left(\dtwo +2-\frac{6m}r\right)^{-1},
\end{equation}
which is not local! However, the action of $\B$ is local with respect to the temporal and radial coordinates. It is only non-local on the surface of the foliating spheres (\ref{foliacja2}).
Because those spheres are compact sets in space-time, the values of $\x$ and $\X$ on any given compact region may depend on some greater subset of space-time, but necessarily still a compact one. We call such an operator \emph{quasi-local}. This is the aforementioned compromise.

Functions $\x$ and $\X$ are conjugate to each other through equations of motion:
\begin{align}
\label{px}
\dot\x&=\frac{f}\Pi \X , \\
\label{pX}
\dot\X&=\frac\Pi{r^2} \left\{\left(f{r^2}\x,{_3}\right),{_3} + \left[\dtwo +f(1-2 \B ) +1 - r^2\Lambda\right] \B \x\right\}.
\end{align}
We may repeat the steps we have taken with the axial invariants and combine these equations into a distorted wave equation. This time, however, the distortion will be a quasi-local operator:
\begin{equation*}
\left\{ \square+\frac{1}{r^2} \left( \dtwo +2-\frac{6m}r\right)^{-2}\left[\dtwo^2\left(\frac{8m}r-\frac 23 r^2 \Lambda \right)
+8\dtwo \left(1-\frac {3m}r \right) \left(\frac mr - \frac{r^2}3 \Lambda \right) + \right.\right. \notag
\end{equation*}
\begin{equation}
+\left. \left. 8 \left(-\frac{2m}r+\frac{3m^2}{r^2}-\frac{r^2}3 \Lambda+2mr\Lambda \right) \right] \right\} \x=0  .
\label{rfx}
\end{equation}
We may recast this equation into a form analogous to (\ref{RWE}), obtaining a rather uninviting expression for the potential:
{\small \begin{equation}
 V^{(+)} := -\frac f{r^2}\left[(\dtwo+2)^2\left(\dtwo -\frac{6m}r\right)+\frac{36m^2}{r^2}\left(\dtwo +2 -\frac{2m}r+\frac 23 r^2 \Lambda\right)\right]
\left(\dtwo+2-\frac{6m}r\right)^{-2} \, .
\end{equation}}%
Finally, splitting $\x$ into spherical harmonics, $\x = \exp (i \sigma t)Y_l(\theta,\phi) Z^{(+)}(r)/r$, and replacing the radial coordinate in the same way as in (\ref{RWEsplit}), we arrive at a generalized version of the well-known Zerilli equation. A polar counterpart to the axial Regge--Wheeler equation:
\begin{equation}
 \left(\frac{d^2}{dr*^2}+\sigma^2 \right)Z^{(+)}Y_l=V^{(+)}Z^{(+)}Y_l.
\end{equation}

\subsection{Mono-dipole part of the invariants}
\label{mono-dipole}
We would like to begin deeper analysis of the invariants by taking a separate look at their monopole and dipole parts in the decomposition with respect to spherical harmonics, as they play a significantly different role then the higher multipoles. We will denote those parts by $\mon(\x)$ and $\dip(\x)$ respectively, and the rest (i.e.``mono-dipole-free" part), which we will call ``radiation part'', by $\underline\x$. It is easy to notice that, by definition of the invariants, $\dip(\x)$ and $\mon(\y)$ vanish identically. We can therefore write:
\begin{align}
\x & = \mon ( \x ) + \underline{ \x },\\
\y & = \dip ( \y ) + \underline {\y}.
\end{align}

The formulae (\ref{Y}) and (\ref{X}) imply vanishing of the whole ``mono-dipole" part of the conjugate invariants $\X$ and $\Y$. Therefore $\mon ( \x )$ and $ \dip ( \y ) $ are constant in time, due to relations (\ref{py}) and (\ref{px}).

The last observation, crucial for the distinction between the mono-dipole and the radiation part, concerns the constraint equations - for higher multipoles they now simply express the relation between our invariants and the redundant parts of metric and momentum, and will be used in section \ref{reconstruction} to reconstruct the original initial data from the invariants. The functions ($\underline{\x}, \underline{\X}, \underline{\y}, \underline{\Y}$) are therefore truly unconstrained in our theory. In the mono-dipole part, however, those ``redundant parts'' vanish and the constraint equations strictly specify the behavior of $\mon(\x)$ and $\dip(\y)$ with respect to the radial coordinate.

Multiplying (\ref{rotww}) by $\frac {2r^2}\Pi$ and taking the dipole part, we obtain:
\begin{equation}
( r^2 \dip(\y))_{,3}=0,
\end{equation}
which can be solved immediately:
\begin{equation}
\dip ( \y ) = \frac \alpha {r^2}.
\end{equation}

The case of $\mon (\x)$ is somewhat harder. Through manipulations on formula (\ref{wsk}) we arrive
at:
\begin{equation}                                                                                                                                                                                                                           -\left( r  \B ^{-1}\x\right)_{,3} = r^2\chi^{AB}{}_{||AB}
-\left(r^3 \B ^{-1}\chi^{AB}{}_{||AB}\right)_{,3} - \dtwo h^3{}_3
+\frac 12 r \dtwo H_{,3}+\left(2-\frac{6m}{r}\right)rh_3{^A}{}_{||A}.
\end{equation}
Restricting this formula to the monopole part kills the right side entirely and turns $\cal B$ into a simple multiplication operator. The solution to the resulting equation is now obvious:
\begin{equation}
-\Big((r-3m)\mon(\x)\Big)_{,3}=0 \quad \Rightarrow \quad \mon(\x)=\frac{\beta}{r-3m}.
\end{equation}

The integration constants $\alpha$ and $\beta$ are ``conserved charges'' and can be interpreted as the angular momentum and mass of the metric perturbation.

\subsubsection{Momentum and center of mass for (Anti)de Sitter}
\label{AdScharge}
Existence of conserved charges corresponds to the symmetries of the background metric. If we increase the number of symmetries, new charges will appear.

We will set $m=0$ in this section, restricting ourselves to an (Anti)de Sitter  background. In this situation we may take $\B$ to be simply an identity operator. The limit of $\B$ for vanishing $m$ is, of course, not equal to $1$ due to the vanishing of the dipole part. However, if we assume $m=0$ from the beginning and rederive the invariants, we will not need to introduce $\B$ at all! What follows --- the dipole part of $\x$ and $\X$ no longer needs to vanish. The radial and temporal dependence of these functions, however, is strictly set by the scalar constraint (\ref{wsk}) and equations of motion. We present here the appropriate derivation.

We begin by finding the radial dependence of $\dip (\x)$ through integration of the scalar constraint. For brevity, from now on until the end of the section we will stop explicitly denoting the $\dip$ operator, and just remember that we deal only with the dipole part of the equations. Let us compare the definition of $\x$ with the constraint equation. We will underline the elements with vanishing dipole part:
\begin{equation}
\begin{aligned}
\x &= \underline{r^2\chi^{AB}{_{||AB}}}-\underline{\frac{1}{2}(\dtwo+2) H}+
   \underset{=1}{\B} \left[ 2 h^{33}+2r h^{3C}{_{||C}} -rf H,{_3}  \right] \\
   &= 2 h^{33}+2r h^{3C}{_{||C}} -rf H,{_3}.
\end{aligned}
\end{equation}
The constraint equation:
\begin{equation}
\frac{\sqrt f}{r^3}\left[ {r^2}{\sqrt f}
 (rH,{_3}-2rh_{3A}{^{||A}}-2h_3{^3})\right],{_3}
 +\underline{\frac 1{r^2}(\dtwo+2)h^3{}_3}
 +\underline{\frac{1}{2r^2}(\dtwo+2)H )}
 - \underline{\chi^{AB}{_{||AB}}} = 0.
\end{equation}
The comparison gives us:
\begin{equation}
\frac{\sqrt{f}}{r^3}\left[\frac{\x r^2}{\sqrt{f}}\right],{_3}=0,
\end{equation}
from which we immediately integrate:
\begin{equation}
\x=\frac{\alpha \sqrt{f}}{r^2}.
\end{equation}
The form of $\X$ quickly follows, as a consequence of (\ref{px}):
\begin{equation}
\frac{\X}{\Pi}=\frac{\dot{\alpha}}{r^2\sqrt{f}}.
\end{equation}

To obtain the temporal dependence, we perform a direct calculation of the right hand side of (\ref{pX}):
\begin{equation*}
\frac{\dot{\X}}{\Pi}=\frac{\alpha\Lambda}{3r^2\sqrt{f}},
\end{equation*}
which leads to a simple differential equation for $\alpha$:
\begin{equation}
\ddot{\alpha}=\alpha\frac \Lambda 3. \label{ppalpha}
\end{equation}
If we denote the initial values of invariants by $\x_0$ and $\X_0$, we can write down the solution for $\x$:
%\begin{equation}
%\x=\frac 12\left(\x_0+\frac{\X_0}{\Pi}\frac{f}{\sqrt{\Lambda/3}}\right)e^{\sqrt{\frac{\Lambda}{3}}t}+
%\frac 12\left(\x_0-\frac{\X_0}{\Pi}\frac{f}{\sqrt{\Lambda/3}}\right)e^{-\sqrt{\frac{\Lambda}{3}}t}.
%\end{equation}
%{\alteration
\begin{equation}
\x= \x_0 \cosh\left(t\sqrt{\frac \Lambda 3}\right)+
\frac{\X_0}{\Pi}\frac{f}{\sqrt{\Lambda/3}} \sinh \left(t\sqrt{\frac \Lambda  3}\right). \label{COM}
\end{equation}
%}
The formula above describes either oscillation or exponential growth/shrinkage, depending on the sign of $\Lambda$. It is noteworthy that in the special case of $\Lambda=0$ (that is --- a Minkowski background), equation (\ref{ppalpha}) gives us the simple solution of linear movement:
\begin{equation}
\x=\x_0+\frac{\X_0}{\Pi}t,
\end{equation}
which suggests an interpretation of the dipole components of $\x$ and $\X$ as the center of mass of the perturbation and its momentum.
%{\alteration
The behavior of equation (\ref{COM}) can be easily understood in this terms --- a repelling (positive) cosmological constant causes the center of mass to be propelled towards the cosmological horizon, while an attractive (negative) one results in an orbit--like evolution.
Unsurprisingly, this mirrors the equations of motion derived form an appropriate non-relativistic limit of the (anti) de Sitter spacetime \cite{newtonhooke}.

All radial functions corresponding to conserved charges exhibit singular behavior at some point ($r=3m$ in case of the mass of the perturbation and at the origin for the rest). This is a known property --- some additional commentary can be found for example in \cite{charges}.
To impose a smoothness condition on the perturbation, one must set the charges to zero. This may be in some cases done by an appropriate modification of the data --- $\mon(\x)$ can be interpreted as a discrepancy between the mass parameter of the background and the total mass of the metric being linearized. A change of $m$ can remove this discrepancy. In a somewhat similar manner --- a nonzero value of $\dip(\x)$ and $\dip(\X)$ suggests a wrong choice of ``frame of reference'', to be fixed by an appropriate translation and boost.

%}

\subsection{The symplectic form}
\label{symplectic}

By taking linear combinations {\alteration of} functions of $(\x,\X,\y,\Y)$ one can easily produce equivalent sets of invariants. There are, however, several reasons for our particular choice. {\alteration The main} justification comes from its relation with the symplectic form of the ADM formulation of linear gravity, $\Omega := \int_\Sigma \delta P^{kl}\wedge \delta h_{kl}$.
This expression is not entirely gauge-independent. However, the gauge action restricts itself to boundary terms:
\begin{equation}
\begin{aligned}
\int_\Sigma \delta P^{kl}\wedge \delta h_{kl} \quad \rightarrow & \quad \int_\Sigma \delta P^{kl}\wedge \delta h_{kl} +2\int_{\partial \Sigma}\delta P^3{}_l\wedge \delta \xi^l \\
&\hspace*{-2cm} +\int_{\partial\Sigma}\sqrt{\eta} \left[ \delta(N\xi^0)_{|k}\wedge \delta h^{3k} - \delta(N\xi^0)^{|3}\wedge \delta h +\delta(N\xi^0)\wedge\delta(h^{|3}-h^{3l}{}_{|l}) \right],
\end{aligned}
\end{equation}
and with further assumption that the gauge field and the normal derivative of its temporal part are fixed on the boundary, $(\delta\xi^\mu |_{\partial\Sigma}=0)$, $(\delta\xi^0{}_{|3}|_{\partial\Sigma}=0)$, it vanishes entirely.

We have already shown that the components of initial data are not {\alteration independent from one another}. This redundancy can be removed through application of the constraint equations and some geometric identities, yielding a reduced version of the symplectic form and making some of its physical properties more apparent. The expression under the integral splits with respect to the decomposition into spherical harmonics, which allows us to separate the monopole and dipole part of the data (which contain the conserved charges) from higher multipoles.
If we now take a look at the mono-dipole free part of the form, which describes the radiation, it readily expresses itself in terms of our invariant quantities:
\begin{equation}
\begin{aligned}
 \int_\Sigma \delta \underline{P}^{kl}\wedge \delta \underline{h}_{kl} =&\int_\Sigma \delta \underline{\X} \wedge \dtwo^{-1}(\dtwo+2)^{-1} \delta\underline{\x} + \delta\underline{\Y}\wedge \dtwo^{-1}(\dtwo+2)^{-1}\delta \underline{\y} \\
 &+\int_{\partial\Sigma} \delta(r\underline{\Xi})\wedge\dtwo^{-1}(\dtwo+2)^{-1}\delta(f\B \underline{Q})-\delta(r\underline{P^{3A}{}_{||A}})\dtwo^{-1}\delta \underline{H}\\
 &+\int_{\partial\Sigma}\delta(r^2\underline{\chi_A{}^B{}_{||BC}\varepsilon^{AC}})\wedge \dtwo^{-1}(\dtwo+2)^{-1}\delta(\Pi\underline{\y}),
\end{aligned}
\end{equation}
where $Q$ and $\Xi$ are auxiliary functions given by the formulae:
\begin{align}
\Xi &:= 2rP^{3A}{}_{||A} +\dtwo P^3{}_3 ,\\
Q &:= 2 h^3{}_3 +2r h_3{}^A{}_{||A}-rH_{,3}.
\end{align}
Some gauge-dependent expressions remain in the boundary terms (which do not affect the dynamics). This is unavoidable --- as we noted before, the whole form is not entirely gauge-invariant.

The remaining mono-dipole part takes the form:
\begin{equation}
\begin{aligned}
{\alteration \Omega-\underline{\Omega}= }&\int_\Sigma \frac 12 \delta P_{33}\wedge\B^{-1}\delta\mon(\x) - \Pi\dip(\delta\y)\wedge\dtwo^{-1}\delta(h_{3A||B}\varepsilon^{AB}) \\
&+\int_{\partial \Sigma} \frac 12 r \delta P^3{}_3\wedge\delta\mon(H) + \frac{r^3N}{12m}(N\delta\Xi)_{,3}\wedge\delta\dip(H)+\frac{r^2 f}{12m}\delta\Xi\wedge\delta\dip(Q).
\end{aligned}
\end{equation}

The derivation of these formulae has been described in detail for Schwarzschild in \cite{praca}, but it carries over to Kottler with hardly any alterations, {\alteration so we do not repeat it here}.
The only significant difference is that we should consider the situation in which $m=0$. In that case the contribution from the dipole polar part is no longer a purely boundary term and instead takes the form:
\begin{equation}
\int_{\Sigma} -\frac 12 \delta\dip(\X)\wedge\delta h^3{}_3+\frac 12 r \delta P_3{}^{A}{}_{||A}\wedge\delta\dip(\x) + \int_{\partial \Sigma}\frac 12 r^2 \delta P^{3A}{}_{||A}\wedge \delta \dip(H).
\end{equation}

The total reduced phase space measure defined by this symplectic form has been investigated in \cite{entropy}, in the case of a Schwarzschild background. The volume of reduced phase space has been shown to diverge to infinity for the region between the black hole horizon and infinity. This result follows from an observation, that the measure is proportional to a divergent integral of the form $\int_{r_0}^{r_\infty}\dd r /{f(r)}$, which remains true for the Kottler metric case. Just as in pure Schwarzschild case, the integral remains divergent in the presence of a cosmological constant both for $r_0$ approaching the Schwarzschild radius and for $r_\infty$ going to infinity or approaching the cosmological horizon, depending on whether we are in the Anti de Sitter or de Sitter case.

{\alteration
\subsection{Energy and momentum of the perturbation}

From the symplectic form we can easily obtain the Hamiltonian of the theory. The $\sim$ symbol here denotes equality up to boundary terms:
\begin{equation}
\begin{aligned}
\underline{\Omega}(\frac{\p}{\p t},\cdot) \sim & \int_\Sigma  \underline{\dot{\X}} \dtwo^{-1}(\dtwo+2)^{-1} \delta\underline{\x} -\underline{\dot{\x}} \dtwo^{-1}(\dtwo+2)^{-1} \delta\underline{\X} + \\
&+\underline{\dot{\Y}} \dtwo^{-1}(\dtwo+2)^{-1}\delta \underline{\y} - \underline{\dot{\y}} \dtwo^{-1}(\dtwo+2)^{-1}\delta \underline{\Y} \\
&= - 16 \pi \delta \mathcal{H} + \int_{\partial \Sigma}\frac{\Pi f}{r}\left[ (r\underline{\y})_{,3} \dtwo^{-1}(\dtwo+2)^{-1} \delta \underline{\y} +(r\underline{\x})_{,3} \dtwo^{-1}(\dtwo+2)^{-1} \delta \underline{\x} \right] 
\end{aligned}
\end{equation}
The Hamiltonian function obtained in this way expresses itself in terms of the invariants through the following integral:
\begin{equation}
\begin{aligned}
16\pi \mathcal{H}&=\frac 12 \int_{\Sigma} \frac{f}{\Pi}\left[ \underline{\X} \dtwo^{-1}(\dtwo+2)^{-1} \underline{\X} + \underline{\Y} \dtwo^{-1}(\dtwo+2)^{-1} \underline{\Y} \right] + \\
&+\frac 12\int_{\Sigma} \frac{\Pi}{r^2} \left[ f(r\underline{\x})_{,3} \dtwo^{-1}(\dtwo+2)^{-1} (r\underline{\x})_{,3} + \underline{\x} \frac{r^2}{f} V^{(+)} \dtwo^{-1}(\dtwo+2)^{-1} \underline{\x} \right] +  \\
&+\frac 12\int_{\Sigma} \frac{\Pi}{r^2} \left[ f(r\underline{\y})_{,3} \dtwo^{-1}(\dtwo+2)^{-1} (r\underline{\y})_{,3} + \underline{\y} \frac{r^2}{f} V^{(-)} \dtwo^{-1}(\dtwo+2)^{-1} \underline{\y} \right],
\end{aligned}
\end{equation}
which gives us a natural candidate for the density of energy carried by the perturbation.
We can immediately check two important properties expected from a reasonable candidate for density of energy of a gravitational field. Firstly,
the expression under the integral is obviously non-local.
Secondly, it is positive definite in the part of spacetime outside the event horizon (where $f>0$) --- this follows from the observation that $V^{(+)}$ and $V^{(-)}$ are both positive definite when $f>0$ and $l\geq 2$.

We refer the reader to \cite{smolka}, where a similar energy functional is considered on Minkowski background, compared to several super-energy functionals and shown to be equal, up to boundary terms, to an energy functional obtained from quantum considerations \cite{birula}:
\begin{equation}
16\pi \overline{\mathcal{H}}=\int_\Sigma \left[ E^{ab}(-\bigtriangleup)^{-1}E_{ab}+B^{ab}(-\bigtriangleup)^{-1}B_{ab}\right],
\end{equation}
where $E$ and $B$ are appropriate parts of the Weyl tensor: $E_kl=W_{0kl0}$, $B_{kl}=\frac 12 \varepsilon_l{}^{ij}W_{0kij}$, and $\bigtriangleup$ is the three-dimensional Laplace operator.

In a similar way, by contracting the symplectic form with the Killing field connected to rotations, we can obtain a candidate for the density of angular momentum. Taking the $\frac{\partial}{\partial \varphi}$ as an example and simplifying through integration by parts:
\begin{equation}
\underline{\Omega}(\frac{\p}{\p \varphi},\cdot) \sim 
 -2\int_\Sigma  \underline{\x}_{,\varphi} \dtwo^{-1}(\dtwo+2)^{-1} \delta\underline{\X} + 
 \underline{\y}_{,\varphi} \dtwo^{-1}(\dtwo+2)^{-1}\delta \underline{\Y} = - \delta \mathcal{J}_z
\end{equation}
\begin{equation}
\mathcal{J}_z = \int_\Sigma  \underline{\x}_{,\varphi} \dtwo^{-1}(\dtwo+2)^{-1} \underline{\X} + 
 \underline{\y}_{,\varphi} \dtwo^{-1}(\dtwo+2)^{-1} \underline{\Y}
\end{equation}

\section{Stationary solutions}
As a part of our analysis of the system we would like to find stationary solutions of equations of motion. We will first do so for the simpler, axial part and then proceed to the polar solutions.

}
\subsection{Stationary solutions for $\y$}

We once again split the invariants into spherical harmonics and denote by $\delta$ the eigenvalue of $\dtwo$ (note that it is a negative number). To simplify the notation, let us also substitute $\lambda:=\frac \Lambda 3$.

If we assume vanishing of the time derivatives, the equations of motion translate into:
\begin{align}
0&=\Y ,\\
0&=\y_{,33}(r^3-2mr^2-r^5\lambda)+\y_{,3}(2r^2-2mr-4r^2\lambda)+\y(\delta r+8m-2r^3\lambda).
\end{align}

To analyze solutions of these equations, we expand $\y$ into a power series in $r$. However, we must remember that our solutions must be well defined only on some interval of the radial coordinate (\ref{foliacja2}), which does not necessarily include $r=0$. We shouldn't therefore \emph{a priori} disregard negative powers of $r$. We postulate a Laurent expansion:
$\y=\sum_{k=-\infty}^\infty y_kr^k$ and obtain the following linear recurrence formula for expansion coefficients:
\begin{equation}
2m\left[4-k^2\right]y_k+\left[k(k-1)+\delta\right]y_{k-1}-\lambda(k-2)(k-1)y_{k-3}=0.
\end{equation}
Properties of the obtained equation depend heavily on the exact values of $m$, $\lambda$ and $\delta$. We will not, therefore, look for one universal solution, but instead try to analyze properties of the expansion on a case-by-case basis.

For further investigation it will be useful to insert the explicit value of $\delta=-l(l+1)$, $l\in \mathbb{N}\backslash \{0\}$. We can then rewrite our formula as:
\begin{equation}
\label{coeff_full}
-2m(k-2)(k+2)y_k+(k+l)(k-l-1)y_{k-1}-\lambda(k-2)(k-1)y_{k-3}=0.
\end{equation}

Now we split the problem into separate cases:
\begin{itemize}
\item The simplest case is of course the Minkowski spacetime: $m=0$, $\lambda=0$. In this situation our formula boils down to:
\begin{equation}
y_k(k+l)(k-l-1)=0,
\end{equation}
which allows exactly two expansion coefficients to be non-zero for each spherical harmonic.
\item For $m \neq 0$, $\lambda = 0$ our formula simplifies significantly:
\begin{equation}
\label{coeff_m}
2m(k-2)(k+2)y_k=(k+l)(k-l-1)y_{k-1},
\end{equation}
but its behavior is not immediately obvious, due to vanishing of the brackets for certain values of $k$. For each spherical harmonic number $l$ the equation allows a two-parameter family of solutions, which turn out to be combinations of two hypergeometric power series:
 $r^2 F(l+3,2-l;5;\frac{r}{2m})$ and $\frac{1}{r^{l+1}}F(l-1,l+3;2l+2;\frac{2m}{r})$. The second one was already inspected in \cite{praca}. It is the only solution regular at infinity, but it diverges logarythmically at the event horizon for $l>2$.

\item For $m = 0$, $\lambda \neq 0$ our recurrence formula turns into:
\begin{equation}
(k+l)(k-l-1)y_{k-1}=\lambda(k-2)(k-1)y_{k-3}.
\end{equation}
By shifting the $k$, we obtain a simpler form:
\begin{equation}
\label{coeff_lambda}
(k+l+1)(k-l)y_{k}=\lambda \, k(k-1)y_{k-2}.
\end{equation}
Let us note that coefficients for even and odd powers of $r$ decouple here.

For each harmonic number the solution turns out to be a three-parameter family, which again can be grouped into a combination of three hypergeometric power series:\\
 \mbox{$r^l F(\frac{l+1}{2},\frac{l+2}{2};l+\frac 32;\lambda r^2)$}, $\frac 1r F(\frac l2,\frac{l+1}{2};\frac 12;\frac{1}{\lambda r^2})$ and $\frac{1}{r^2} F(-\frac{l-1}{2},\frac{l+2}{2};\frac 32;\frac{1}{\lambda r^2})$. It is noteworthy that for each value of $l$ one of the two series in negative powers of $r$ will always be infinite, while the other will reduce to a polynomial of degree $l+1$ in $\frac 1r$.

\item When both $m$ and $\lambda$ are not vanishing, we have to take into account the whole formula (\ref{coeff_full}), being now a recurrence of a third degree. Although we are unable to give explicit formulae for the solutions here, we can still classify them in terms of their expansions, which give us a general idea of their properties.

A following observation proves helpful in the analysis: if we look at (\ref{coeff_full}) for $k\in \{3,4,5\}$, we obtain a set of equations which defines a map: $\mathbb{R}^3 \ni (y_0,y_1,y_2) \to (y_3,y_4,y_5)\in\mathbb{R}^3$. This map is easily seen to be linear and isomorphic for any value of $l$ and any non-vanishing $m$ and $\lambda$. This stays true if we consider any triple of equations for three consecutive values of $k\geq 3$. In an analogous way, the triple of equations for $k\in\{-3,-4,-5\}$ defines an isomorphism between the values of $(y_{-3},y_{-4},y_{-5})$ and $(y_{-6},y_{-7},y_{-8})$ and similar relations exist for all lower $k$'s. This means that we only need to analyze the behavior of eight coefficients $y_k$ for $k\in [-5,2]$ governed by five equations (\ref{coeff_full}) for $k\in[-2,2]$. All the other coefficients can be computed from this set. Moreover, if any of the coefficients from the triple $(y_0,y_1,y_2)$ or $(y_{-3},y_{-4},y_{-5})$ are non-zero, it implies that the expansion is infinite in either positive or negative powers of $r$.

In the mono-dipole free part ($l\geq 2$), the situation turns out to be rather simple. Equations imply $y_1=y_0 \equiv 0$ and two families of solutions exist: one being an infinite series in positive powers of $r$, beginning at $y_2$ and parametrized by its value, and a two-parameter family of infinite expansions in negative powers, parametrized by $y_{-1}$ and $y_{-2}$.

This picture looks somewhat differently for the monopole and dipole part. The space of solutions is now spanned by four functions. One generated by $y_2$ and having an infinite expansion in positive powers of $r$, one generated by $y_0$, with infinite expansion in both directions, one with an infinite expansion in the negative powers of $r$, generated by $y_{-1}$ in the dipole case and by $(3my_{-1}+y_{-2})$ for the monopole, and a solution with finite expansion: $\frac{r-3m}{r^2}$ for the monopole and $\frac{1}{r^2}$ for the dipole.

\end{itemize}

An observant reader will notice that we have obtained several solutions in the mono-dipole part, instead of just two described in \ref{mono-dipole}. This is due to the fact that we have been looking solely at the equations of motion, without considering the restrictions resulting from the definition of $\y$ and the constraints.

\subsection{Stationary solutions for $\x$}

An analogous approach with expanding $\x$ into a Laurent series $\x=\sum_{k=-\infty}^\infty x_kr^k$ applied to equation (\ref{pX}) yields a rather complicated recurrence relation between coefficients of the expansion:
\begin{equation}
\label{xrecurrnece}
\begin{aligned}
0 =& -72m^3(k+1)^2x_{k+1} + 12m^2 \left[ (2\delta+7)k^2+3k +(\delta+2) \right] x_k  \\
&-2m(\delta+2)\left[ (\delta +8)k^2 -2(\delta+5)k + 3(\delta +2) \right]x_{k-1}  \\
&+\left[(\delta+2)^2 (k-l-2)(k+l-1) - 36m^2\lambda(k-2)(k+1)\right]x_{k-2}  \\
&+12m\lambda(\delta+2)[k^2-3k+1]x_{k-3} - \lambda(\delta+2)^2(k-4)(k-1)x_{k-4}.
\end{aligned}
\end{equation}
We will again perform a separate analysis for various cases of $m$ and $\lambda$:
\begin{itemize}
\item The vanishing of mass and cosmological constant in Minkowski spacetime turns (\ref{xrecurrnece}) into a simple equation, similar to the one for $\y$:
\begin{equation}
x_k(\delta+2)^2(k-l)(k+l+1)=0.
\end{equation}
One should notice, however, that the factor $(\delta+2)^2$ is just a remnant of removing the quotient in $\B$ and should be omitted, as was briefly discussed in \ref{AdScharge}.
Therefore the correct equation has the form:
\begin{equation}
x_k(k-l)(k+l+1)=0,
\end{equation}
and yields exactly two non-vanishing expansion coefficients for every multipole.

\item When $m=0 \neq \lambda$ we again can ignore the common factor $(\delta+2)^2$ and obtain:
\begin{equation}
x_k(k-l)(k+l+1)=x_{k-2}\lambda(k-2)(k+1).
\end{equation}
This result seems similar to the analogous case for $\y$, but the shift in factors results in a slight change in solutions and an anomaly for the monopole part. Again, even and odd powers of $r$ decouple in this equation and the solutions can be expressed as a combination of three hypergeometric series. For the harmonic number $l\geq 1$ these series are: \mbox{$r^l F(\frac l2, \frac{l+3}{2};l+\frac 32; \lambda r^2)$}, $F(\frac l2, -\frac{l+1}{2};-\frac 12; \frac{1}{\lambda r^2})$ and $\frac{1}{r^3} F(\frac{l+3}{2},-\frac{l-2}{2};\frac 52; \frac{1}{\lambda r^2})$. As in the analogous case for $\y$, one of the series in negative powers of $r$ will terminate and become a polynomial of degree $l+1$ in $\frac 1r$.

In case of the monopole harmonic the first two solutions become the same (a constant function) and an additional solution appears, which can also be described in terms of a hypergeometric series: $\frac 1r F(1,-\frac 12;\frac 12;\lambda r^2)$.

\item For $\lambda=0$ and $m\neq0$ equation (\ref{xrecurrnece}) becomes a recurrence of a third degree (except for the dipole harmonic):
\begin{equation}
\begin{aligned}
\label{xrec_lam_zero}
0 =& -72m^3(k+1)^2x_{k+1} + 12m^2 \left[ (2\delta+7)k^2+3k +(\delta+2)  \right] x_k  \\
&-2m(\delta+2)\left[ (\delta +8)k^2 -2(\delta+5)k + 3(\delta +2) \right]x_{k-1}  \\
&+(\delta+2)^2 (k-l-2)(k+l-1) x_{k-2}.
\end{aligned}
\end{equation}
We will analyze its properties by treating it as a map between $\{ x_k,x_{k-1},x_{k-2} \}$ and $\{ x_{k+1},x_{k},x_{k-1} \}$. By looking at its matrix representation, it is easy to notice that the isomorphicity of this map depends only on the coefficients next to $x_{k+1}$ and $x_{k-2}$ in equation (\ref{xrec_lam_zero}). The mapping is therefore isomorphic if the following expressions are non-zero: $k+1$, $(k-l-2)(k+l-1)$. Inspection of irregularities of equation (\ref{xrec_lam_zero}) for $k\in \{-1,l+2,-(l-1)\}$ is therefore the key to understanding the set of solutions.

%Another useful observation concerns coefficients next to $x_k$ and $x_{k-1}$ in equation (\ref{xrec_lam_zero}). If we treat them as polynomials in $k$ we can see that they do not posses roots in integer numbers, with two exceptions: for $l=1$ both polynomials posses obvious integer roots and for $l=2$ the coefficient next to $x_{k-1}$ vanishes for $k=2\vee -3$.

The monopole case possesses three separate solutions, which can be best parametrized by values of coefficients $x_0$, $x_{-1}$ and the combination $3mx_1-x_0$. The first parameter governs a solution in positive powers of $r$, the second - a solution in negative powers and the third parameterizes a solution extending in both negative and positive powers.

In the case of the dipole harmonic equation (\ref{xrec_lam_zero}) reduces to a recurrence equation of the first degree:
\begin{equation}
72m^3(k+1)^2x_{k+1}=36m^2 k (k+1) x_k,
\end{equation}
which possesses exactly two solutions: a constant function and $\ln(1-\frac{2m}{r})$, easily found by inspecting the original differential equation.

The situation becomes more complicated for higher multipoles. For $l\geq 2$ the family of solutions is three-dimensional and we may choose $x_0$, $x_{-1}$ and $x_{-(l+1)}$ as parameters. $x_{-(l+1)}$ governs a solution extending infinitely into negative powers of $r$. The exact form of the solutions governed by $x_0$ and $x_{-1}$, however, remains unclear.

%It can be easily seen that all coefficients from $x_{-1}$ down to $x_{-l}$ are pairwise linearly dependent.

Direct inspection of initial values of $l$ (which we performed up to $l=7$) shows a surprising pattern: $x_{l+1}$ and $x_{l+2}$ (which fully determine all higher powers of $r$, due to degeneration of (\ref{xrec_lam_zero}) for $k=l+2$) turn out to be linearly dependent and can be defined in terms of a certain combination of $x_0$ and $x_{-1}$ (different for each $l$). In such case we can choose a following pair of solutions to complete the set: one beginning from $x_0$ and extending infinitely into positive powers of $r$ and a second one, parametrized by the appropriate combination of $x_0$ and $x_{-1}$, which goes in positive powers only up to $x_l$, but extends infinitely into negative powers of $r$.
We have not been able to determine whether the linear dependency of $x_{l+1}$ and $x_{l+2}$ (and with it, our proposed classification of solutions) holds for all values of $l$, because the number of equations that must be taken into consideration grows with $l$.

\item When neither $\lambda$, nor $m$ vanish we have to deal with the full equation (\ref{xrec_lam_zero}), a recurrence of the fifth degree. We again apply the method of treating the equation as a map between consecutive sets of coefficients. Surprisingly, finding a coarse description of the solutions turns out to be much easier than in the previous case, as the placement of irregularities does not depend on $l$ and the map degenerates simply for $k\in \{4,1,-1\}$.

For $l\neq 2$ all solutions can be parametrized by $(x_1,x_0,x_{-1},x_{-2},x_{-3})$. All solutions that extend into infinite positive powers of $r$ are governed by the first four coefficients in this list. The solutions extending into negative powers are determined by the last three coefficients and the combination $\left[-6m x_1+(\delta+2)x_0\right]$.

In the dipole case one can return to the original differential equation, which now becomes relatively simple: $(fr^2\x_{,3})_{,3}=0$. The solutions are given by an integral $\int\frac{\dd[r]}{fr^2}$, which can be computed through partial fraction decomposition.

\end{itemize}

Again, unphysical solutions appear in the mono dipole part, which are excluded by the constraint equations and the definition of $\x$.
%{\alteration
Some commentary is also in order concerning the physical significance of the stationary solutions for higher multipoles. Some of them can obviously be excluded by appropriate conditions imposed on their behavior near the origin or at infinity. We believe that the rest may correspond to linearizations of some stationary solutions to the full Einstein equations. This should mirror in some way the correspondence of the conserved charge $\dip(\y)$ to the angular momentum of a perturbation created by linearizing a Kerr-de Sitter metric on a Kottler background. 
%}

\section{Reconstruction of initial data from the invariants}
\label{reconstruction}
In applications of the invariant formalism proposed in this paper, it may be necessary to invert the transformation between initial data and the invariants. We present a method of reconstructing the perturbation of ADM data with help of the gauge transformations. By doing so, we also prove that our construction of invariants does not lose any physical information.

This section follows closely a similar reasoning presented in \cite{praca}, with small changes to the formulae due to the presence of the cosmological constant and some minor corrections.

Several preparatory steps are in order. We will again make use of the auxiliary quantities introduced in section \ref{symplectic}:
\begin{align}
\Xi &:= 2rP^{3A}{}_{||A} +\dtwo P^3{}_3 ,\label{Xi_def}\\
Q &:= 2 h^3{}_3 +2r h_3{}^A{}_{||A}-rH_{,3}. \label{Q_def}
\end{align}
Rewriting scalar and vector constraints in terms of these variables yields:
\begin{align}
\frac{\sqrt{f}}{r}(r^2\sqrt{f}Q)_{,3} + r^2\chi^{AB}{}_{||AB} - \frac 12 (\dtwo+2)H - \left(\dtwo+2-\frac{6m}{r}\right)h^3{}_3=0, \label{scalar_constraint_Q}\\
r\sqrt{f}(\sqrt{f}\Xi)_{,3}+\left(\dtwo+2-\frac{6m}{r}\right)r P^{3A}{}_{||A} + 2fr^2S^{AB}{}_{||AB}=0.
\label{vector_constraint_Xi}
\end{align}
It will also be helpful to know the action of gauge transformation on $Q$:
\begin{equation}
Q \quad\to\quad Q + \frac 2r \left( \dtwo +2 -\frac{6m}{r}\right)\xi_3,
\end{equation}
and the evolution equation for $\Xi$:
\begin{equation}
\begin{aligned}
\frac{1}{\Pi}\dot{\Xi}=&\frac{1}{2r^2}\dtwo\left(\dtwo+2-\frac{6m}{r}\right)(h^3{}_3-h^0{}_0)+\frac fr (\dtwo+2)h_{3A}{}^{||A} \\
&- \frac fr (r^2\chi^{AB}{}_{||AB})_{,3} -\frac{f}{2r^3}\dtwo(r^2Q)_{,3}.
\end{aligned}
\end{equation}

We will now proceed to recover the twelve component functions defined in section \ref{DataSplitting}, from which the tensors $h_{kl}$ and $P^{kl}$ can be reconstructed directly. Through equations of motion we can also recover components $h^0{}_0$ and $h^0{}_k$, corresponding to the perturbation of shift and lapse, and obtain the full metric perturbation $h_{\mu\nu}$.
As the reader might already expect --- monopole and dipole degrees of freedom require a different approach from the higher multipoles and will be examined separately.

\subsection{Dipole polar part}

In this section only dipole parts of the variables are considered and we denote them with the same symbols as the full quantities. As some of the degrees of freedom of initial data vanish identically in the dipole part, the only quantities we need to reconstruct are $h^3{}_3$, $H$ and $h_3{}^A{}_{||A}$ for the metric perturbation and $P^3{}_3$, $S$ and $P^{3A}{}_{||A}$ for the ADM momentum.

\subsubsection{Momentum components}

From various parts of the vector constraint, we obtain:
\begin{align}
rP^{3A}{}_{||A}&=\frac{r^2\sqrt{f}}{6m}(\sqrt{f}\Xi)_{,3} ,\\
P^3{}_3 &= \frac{r^2\sqrt{f}}{6m}(\sqrt{f}\Xi)_{,3} - \frac 12 \Xi ,\\
S &=\left[ \frac{r^3\sqrt{f}}{6m}(\sqrt{f}\Xi)_{,3}\right]_{,3}.
\end{align}
The whole information about the dipole polar part of $P^{kl}$ is therefore encoded in $\Xi$.

\subsubsection{Metric components}

From the scalar constraint we obtain:
\begin{align}
h^3{}_3 & =-\frac{\sqrt{f}}{6m}(r^2\sqrt{f}Q)_{,3},\\
2rh_3{}^A{}_{||A} & = Q + \frac{\sqrt{f}}{3m}(r^2\sqrt{f}Q)_{,3}+rH_{,3}.
\end{align}
Knowledge of $Q$ and $H$ is therefore sufficient to reconstruct the dipole polar part of $h_{kl}$.

\subsubsection{Dipole polar gauge}

The gauge acts on the dipole part of the quantities $\Xi$, $Q$ and $H$ in the following way:
\begin{align}
-\frac{r^3}{12m\Pi}\Xi &\quad\to\quad -\frac{r^3}{12m\Pi}\Xi + \xi^0 ,\\
-\frac{r^2 f}{12m}Q &\quad\to\quad -\frac{r^2 f}{12m}Q + \xi^3 ,\\
\left(\frac 12 H + \frac{rf}{6m}Q \right) &\quad\to\quad \left(\frac 12 H + \frac{rf}{6m}Q \right) + \xi^A{}_{||A}.
\end{align}
One can see, that by performing a quasi-local gauge transformation one can always set $\Xi$, $Q$ and $H$ to zero.

\subsubsection{Lapse and shift components}

In the dipole polar part, the lapse and shift components are given by appropriate evolution equations:
\begin{align}
-h^0{}_0 & =\frac{r^3}{6m\Pi}\dot{\Xi}+\left(\frac 16 - \frac{r^3\Lambda}{18m}\right)Q, \\
h_{03}   & =\frac{r^2 f}{12 m \Pi} (r\Xi)_{,3} - \frac{r^2}{12m}\dot{Q}, \\
h_0{}^A{}_{||A} & = \frac 12 \dot{H}+\frac{rf}{6m}\dot{Q} - \frac{rf}{6m\Pi}\Xi \, .
\end{align}
We have therefore shown that the whole dipole polar part of $h_{\mu\nu}$ and $P^{kl}$ is just an artifact of a quasi-local gauge freedom, as long as the mass parameter $m$ of the background metric is non-zero.

\subsubsection{The case of vanishing mass parameter $m=0$}

The situation changes significantly, when the mass parameter of the background vanishes. Recall from section \ref{AdScharge} that in that case the dipole polar part of initial data contains a conserved charge and therefore cannot be just a remnant of the gauge. From (\ref{x}), (\ref{X}), (\ref{Q_def}) and (\ref{Xi_def}) one can immediately see:
\begin{align*}
\x&=fQ ,\\
\X&=\Xi.
\end{align*}
To reconstruct particular components of the metric and momentum, we need to introduce some gauge conditions. A rather simple choice is:
\begin{equation}
h^3{}_3=H=0 ,\qquad P^{3}_3 =0.
\end{equation}
This can be realized by obtaining functions $\xi^3$, $\xi_A{}^{||A}$ and $\xi^0$ from integration of appropriate gauge transformations --- (\ref{hxi33}), (\ref{pxi33}) and
\begin{equation}
H \quad\rightarrow\quad H + 2\xi_A{}^{||A}+\frac{4}{r}\xi^3.
\end{equation}
We are then left with:
\begin{align}
h_3{}^A{}_{||A}&=\frac{\x}{2rf},\\
P^{3A}{}_{||A}&= \frac{\Xi}{2r} ,\\
S&=\frac 12 \Xi_{,3},
\end{align}
where the last equality is a consequence of (\ref{wwA}). Values of $h_{03}$, $h_0{}^A{}_{||A}$ and $h^0{}_0$ can then be integrated from (\ref{h33}), (\ref{h3A}) and (\ref{P33}).

\subsection{Radiation polar part, Regge--Wheeler gauge}

We are now dealing with the mono-dipole-free polar part of the variables. There are eight components we need to reconstruct: $\chi^{AB}{}_{||AB}$, $h_{3A}{}^{||A}$, $h^3{}_3$, $H$, $S^{AB}{}_{||AB}$, $P^3{}_3$, $S$ and $P^{3A}{}_{||A}$, plus the lapse and shift. To this end, we impose the following set of quasi-local gauge conditions:
\begin{align}
\chi^{AB}{}_{||AB}=h_{0A}{}^{||A}=h_{3A}{}^{||A}=0,
\end{align}
which is possible due to the following form of gauge transformations:
\begin{align}
r^2\chi^{AB}{}_{||AB} &\quad\to\quad r^2\chi^{AB}{}_{||AB}+ (\dtwo+2)\xi^A{}_{||A} ,\\
r^2h_{0A}{}^{||A} &\quad\to\quad r^2h_{0A}{}^{||A} + \dtwo \xi_0 +r^2\dot{\xi}^A{}_{||A} ,\\
r^2h_{3A}{}^{||A} &\quad\to\quad r^2h_{3A}{}^{||A} + \dtwo \xi_3 +r^2(\xi^A{}_{||A})_{,3}. \label{regge-wheeler gauge}
\end{align}

\subsubsection{Momentum components}

From the evolution equations we obtain:
\begin{equation}
\frac{2f}{\Pi}S^{AB}{}_{||AB}=\dot{\chi}^{AB}{}_{||AB}-\frac{1}{r^2}(\dtwo+2) h_{0A}{}^{||A},
\end{equation}
which yields $S^{AB}{}_{||AB}=0$. The variable $\Xi$ is then immediately given by the definition of $\X$:
\begin{equation}
\label{X_bis}
\X=2r^2 S^{AB}{}_{||AB}+\B \Xi.
\end{equation}
We may then reconstruct other momentum components from the definition of $\Xi$ and appropriate components of the vector constraint (\ref{Xi_def}), (\ref{vector_constraint_Xi}), (\ref{ww3}).

\subsubsection{Metric components}
The definition of $\x$:
\begin{equation}
\label{x_bis}
\x=r^2\chi^{AB}{}_{||AB}-\frac 12 (\dtwo+2)H+f\B Q,
\end{equation}
along with the definition of $Q$ (\ref{Q_def}) and the scalar constraint (\ref{scalar_constraint_Q}) forms a system of quasi-local equations, from which we may obtain the values of $Q$, $H$ and $h^3{}_3$, reconstructing the remaining metric components:
\begin{align}
\frac 12 \dtwo Q     & = \left(r\B^{-1}\x\right)_{,3} ,\\
\frac 12 (\dtwo+2)H  & = f\B Q - \x ,\\
2h^3{}_3             & = Q+rH_{,3}.
\end{align}

\subsubsection{Lapse and shift components}
The remaining lapse and shift components can be computed from equations of motion --- $h_{03}$ from the divergence of (\ref{h3A}):
\begin{equation}
2r^2\Pi^{-1}P^{3A}{}_{||A}+\dtwo h_{03}=r^2\dot{h}_{3A}{}^{||A}-r^2(h_{0A}{}^{||A})_{,3},
\end{equation}
and $h^0{}_0$ from the double divergence of (\ref{SAB}):
\begin{eqnarray}
\frac{2}{\Pi}\dot{S}_{AB}{}^{||AB} &=& \nonumber
\frac{1}{2r^4}\dtwo(\dtwo+2)(h^0{}_0+h^3{}_3)-\frac{1}{r^4}\left[r^2(\dtwo+2)h^{3C}{}_{||C}\right]_{,3}
\\ & &
+
\frac{1}{r^4}\left[(r^2\chi_{AB}{}^{||AB})_{,3}fr^2\right]_{,3}-2\Lambda\chi_{AB}{}^{||AB}.
\end{eqnarray}

\subsection{Radiation polar part. Quasi-local gauge}
\label{quasi-local}
We would like to point out that the Regge--Wheeler gauge is not entirely local in $r$ --- it requires knowledge of data over some interval in the radial variable to compute the radial derivative in the gauge transformation (\ref{regge-wheeler gauge}).
An alternate set of gauge conditions can be considered:
\begin{equation}
Q=H=\Xi=0,
\end{equation}
which is truly quasi-local --- it can be calculated for a single sphere,
as $Q$, $H$ and $\Xi$ transform in the following way:
\begin{align}
rfQ & \quad \to \quad rfQ + 2\left(\dtwo + 2 - \frac{6m}{r} \right) \xi^3 ,\\
\frac{r^2}{\Pi}\Xi & \quad \to \quad \frac{r^2}{\Pi}\Xi -\frac 12 \dtwo  \left(\dtwo + 2 - \frac{6m}{r} \right) \xi^0 ,\\
\hspace*{-1cm} fQ - \frac  12 \left(\dtwo + 2 - \frac{6m}{r} \right)H & \quad \to \quad fQ - \frac  12 \left(\dtwo + 2 - \frac{6m}{r} \right)H -\left(\dtwo + 2 - \frac{6m}{r} \right)\xi^A{}_{||A}.
\end{align}
This yields a different set of equations from which metric and momentum components can be recovered. It is also noteworthy that no gauge conditions are imposed on the components of lapse and shift in this case.

\subsubsection{Metric components}
From the definition of $\x$ (\ref{x_bis}), $Q$ (\ref{Q_def}) and the scalar constraint (\ref{scalar_constraint_Q}) we immediately obtain:
\begin{align}
r^2 \chi^{AB}{}_{||AB} \qquad & = \quad \x, \\
h^3{}_3  = -rh_{3A}{}^{||A} & = \left(\dtwo + 2 - \frac{6m}{r} \right)^{-1}\x.
\end{align}

\subsubsection{Momentum components}
Definition of $\X$ (\ref{X_bis}), $\Xi$ (\ref{Xi_def}) and the vector constraint in form of (\ref{vector_constraint_Xi}) and (\ref{wwA}) allow us to derive:
\begin{align}
2r^2 S^{AB}{}_{||AB} & = \X, \\
S  & =\left[ 2fr\dtwo^{-1}\left(\dtwo + 2 - \frac{6m}{r} \right)^{-1} \X \right]_{,3} - \dtwo^{-1}\X ,\\
\dtwo P^3{}_3  = -2rP^{3A}{}_{||A} & = 2f \left(\dtwo + 2 - \frac{6m}{r} \right)^{-1}\X.
\end{align}

\subsubsection{Lapse and shift components}
Evolution equations yield following relations:
\begin{align}
r\Xi - r\Pi \dot{Q}  & =\left(\dtwo + 2 - \frac{6m}{r} \right)(rP^3{}_3-2\Pi h_{03}) ,\\
f\dot{Q} - \frac 12 \left(\dtwo + 2 - \frac{6m}{r} \right)\dot{H}  & = \frac{f}{\Pi}\Xi - \left(\dtwo + 2 - \frac{6m}{r} \right)h_0{}^A{}_{||A},
\end{align}
which in turn give the following values of lapse and shift:
\begin{align}
h_{0A}{}^{||A}     & = 0 ,\\
h_{03}    \quad    & = \frac{rf}{\Pi}\dtwo^{-1}\left(\dtwo + 2 - \frac{6m}{r} \right)^{-1}\X ,\\
h^{0}{}_{0} \quad  & = \dtwo^{-1}\left(\dtwo + 2 - \frac{6m}{r} \right)^{-1}\left(\dtwo \x - 2f\B \x - 2rf\x_{,3}\right).
\end{align}

\subsection{Monopole polar part}

There are just four components here to reconstruct: $h^3{}_3$, $H$, $P^3{}_3$ and $S$.
We introduce an auxiliary variable $\kappa=\B^{-1}\x$, to get rid of singular behavior of $\mon(\x)$ at $r=3m$. The new variable fulfills the equation:
\begin{equation}
\mon(\kappa)=\mon(\B^{-1}\x) = \left(1-\frac{3m}{r}\right)\cdot\frac{p_0}{r-3m}=\frac{p_0}{r},
\end{equation}
where $p_0$ is an integration constant --- the value of the conserved charge.

\subsubsection{Metric components}
By performing a gauge transformation with an appropriate value of $\mon(\xi^3)$ we can set $H$ to be equal to zero. We then obtain:
\begin{equation}
\mon(h^3{}_3) = \frac{1}{2f} \mon(\kappa)= \frac{p_0}{2rf}.
\end{equation}

\subsubsection{Momentum components}

The gauge transformation of $P^3{}_3$ has the form:
\begin{equation}
\frac{1}{\Pi} P^3{}_3 \quad\to\quad \frac{1}{\Pi} P^3{}_3 -\frac{1}{r^2}\dtwo\xi^0-\frac{2}{r}\xi^{0,3}.
\end{equation}
The monopole part of this gauge transformation gives us an equation for $\xi^0$ (requiring a choice of boundary value), which allows us to set $\mon(P^3{}_3)=0$. From the vector constraint (\ref{ww3}) we then obtain:
\begin{equation}
\mon(S)=0.
\end{equation}
The monopole part of the momentum is therefore only a remnant of the gauge transformation.

\subsubsection{Lapse and shift components}
From the trace of (\ref{hAB}) we obtain:
\begin{equation}
\dot{H}=\frac{2}{\Pi}P^{33}+2h_{0A}{}^{||A}+\frac{4f}{r}h_{03},
\end{equation}
which yields $\mon(h_{03})=0$. From another evolution equation (\ref{P33}) we get:
\begin{equation}
\mon\left(\frac{r}{\Pi}\dot{P}^3{}_3\right) = \mon(-f(h^0{}_0+h^3{}_3)_{,3}),
\end{equation}
which in turn gives us the following solution for $h^0{}_0$:
\begin{equation}
\mon(h^0{}_0)= - \frac{p_0}{2rf}+C(t),
\end{equation}
where $C(t)$ is an integration constant, dependent only on the time coordinate. It can probably be set to 0 with an appropriate choice of the gauge.

\subsection{Axial part of the initial data}
\subsubsection{Momentum components}
The definition of $\y$ and the vector constraint (\ref{wwA}) immediately give us the values of momentum components $P^{3A||B}\varepsilon_{AB}$ and $S^{AB}{}_{||BC}\varepsilon_{A}{}^{C}$.

\subsubsection{Radiation part of the full metric perturbation}
We impose a quasi-local gauge condition:
\begin{equation}
\chi^{AB}{}_{||BC}\varepsilon_{A}{}^{C}=0 , \qquad r^2\chi^{AB}{}_{||BC}\varepsilon_{A}{}^{C} \quad\to \quad r^2\chi^{AB}{}_{||BC}\varepsilon_{A}{}^{C} + (\dtwo +2)\xi_{A||B}\varepsilon^{AB},
\end{equation}
and extract $h_{3A||B}\varepsilon^{AB}$ from the definition of $\Y$. The shift element is, as usual, obtained from the evolution equation:
\begin{equation}
r^2 \dot{\chi}^{AB}{}_{||BC}\varepsilon_{A}{}^{C} = 2\frac{fr^2}{\Pi}S^{AB}{}_{||BC}\varepsilon_{A}{}^{C} + (\dtwo+2)h_{0A||B}\varepsilon^{AB},
\end{equation}
which gives $h_{0A||B}\varepsilon^{AB}=\frac{f}{r^2}(r^2 \y)_{,3}$.

\subsubsection{Mono-dipole part}
The monopole axial part of the initial data vanishes identically. Therefore we need only concern ourselves with the dipole part. We fix a gauge condition:
\begin{equation}
h_{3A||B}\varepsilon^{AB}=0 , \qquad  h_{3A||B}\varepsilon^{AB} \quad\to\quad h_{3A||B}\varepsilon^{AB} + (\xi_{A||B}\varepsilon^{AB})_{,3} ,
\end{equation}
which requires a choice of boundary value. $h_{0A||B}\varepsilon^{AB}$ is then directly bound to the stationary solution of $\dip(\y)$:
\begin{equation}
r^2 \dot{h}_{3A||B}\varepsilon^{AB}=r^2 h_{0A||B}\varepsilon^{AB}+\y.
\end{equation}

{\alteration
\section{Conclusions}

\subsection{Comparison of the sets of gauge invariants}

Linearized gravitation on Schwarzschild and Schwarzschild-de Sitter backgrounds has been very actively investigated in recent years. We would therefore like to compare our approach with several parallel results, especially those of Dotti \cite{nonmodal1}, \cite{nonmodal2} and Chaverra et al. \cite{COS}, as they also pursued to describe the perturbation without preliminary decomposition into spherical harmonics.
The discussion in section \ref{DataSplitting} shows that the structure of the theory provides several guidelines for construction of invariant quantities, which explains strong convergence between independent studies. This is especially visible in the odd part of the perturbation (the splitting of perturbation data into odd and even parts is a practically ubiquitous step in literature), where there are few degrees of freedom to begin with. Perhaps the most surprising coincidence here lies in the form of one of the invariants produced by Dotti through linearization of the contractions of the Weyl tensor. An entirely different way of reasoning which leads to a familiar result:
\begin{equation}
G_- :=\delta\left(\frac{1}{48}W^{\alpha\beta\gamma\delta}W^*_{\alpha\beta\gamma\delta}\right)= \frac{3m}{r^5}\y.
\end{equation}
What follows, the master equation (113) in \cite{nonmodal2}, used to prove the stability of the odd sector, is directly equal to (\ref{RWE}), although our version does not trivialize for $m=0$.

As usual, matters complicate themselves in the even sector. Still, the second invariant of Dotti can be expressed through one of our invariants in a relatively straightforward way:
\begin{equation}
\begin{aligned}
G_+:&=\delta\left((9m-4r+\Lambda r^3)\frac{1}{48}W^{\alpha\beta\gamma\delta}W_{\alpha\beta\gamma\delta}+3r^3 \frac{1}{720}W^{\alpha\beta\gamma\delta;\epsilon}W_{\alpha\beta\gamma\delta;\epsilon} \right) \\
&=\frac{m}{2r^4}\left\{ f(r\x)_{,3}+\left[ \frac{2m}{r^2}(3m-\Lambda r^3)(\dtwo+2)^{-1}+\left(\frac{r-3m}{r}\right)\right]\B\x \right\},
\end{aligned}
\end{equation}
at least in the mono-dipole free part (in the monopole part the formula seems to produce a sign discrepancy), which shows an interesting compliance with the splitting of ``positions'' and ``momenta''. Although we have argued that such splitting is in some way ``natural'', it is not always followed. A noteworthy example can be seen in \cite{kodama}, where the obtained master scalars can be easily seen to mix $\x$ with $\X$ and $\y$ with $\Y$.

A useful key for comparing results in the troublesome even sector can be easily obtained through the covariant generalization of the Zerilli--Moncrief function $\Psi$ (employed e.g. in \cite{nonmodal2} and \cite{COS} mentioned above), which translates directly to our invariant $\x$:
\begin{equation}
\dtwo(\dtwo+2)\Psi=r\x.
\end{equation}
With this one can, for example, retrace the derivation of Chandrasekhar's duality in \cite{nonmodal2} in our framework and observe that the bijection between solutions of Zerilli and Regge--Wheeler equations also respects the splitting of ``positions'' and ``momenta''.

Another interesting lesson may be drawn from comparison of our postulated energy functional with the abstract ``conserved energy'' formula (223) from \cite{nonmodal2}, which was used in the stability proof. The expressions differ by a factor of
$\dtwo(\dtwo+2)$ ! This discrepancy does not, however, influence the validity of the stability proof or its applicability to our set of invariants. The action of operator $\dtwo(\dtwo+2)$ simply multiplies each harmonic mode by a constant factor, which does not influence the dynamics. This shows that to distinguish a correct formula for ``physical energy'' in linearized gravity one cannot depend solely on the evolution of separate harmonic modes --- one needs to find appropriate summation coefficients.

\subsection{Final remarks}
}

We have presented a formalism for description of linear perturbations of the Kottler metric in terms of four gauge-invariant scalar functions. Our construction does not require a choice of a gauge or an \emph{a priori} splitting of the perturbation data into spherical harmonics. {\alteration Also, no potential functions are introduced in this scheme --- all calculations are performed directly on the perturbation components, which makes their interpretation clear and straightforward.}  
The equations of motion for the obtained functions are quasi-local and correspond in an obvious way to the Regge--Wheeler and Zerilli equations. Our particular choice of the four invariant functions is dictated by the symplectic form of the ADM theory 
{\alteration and produces elegant formulae for the density of energy and angular momentum of the dynamic part of the perturbed data. It can be shown that our proposed expression for energy density, with appropriate treatment of boundary terms, is equal to a linearization of certain quasi-local mass candidates in the full theory. We hope to analyze this correspondence further in the future.}

A thorough analysis of the non-dynamical mono-dipole part of the perturbation was performed {\alteration within our} framework. It was shown that the information about the mass and angular momentum of the perturbation is encoded there, and in case of vanishing mass of the background black hole --- position and momentum of the center of mass of the perturbation can also be defined.

We have also presented a full classification of stationary solutions to equations of motion, as this topic seems to be neglected in literature. 
{\alteration 
Unfortunately, many of these solutions are available to us only in terms of the coefficients of their expansion into power series and a further investigation is necessary to determine their properties: can some or all of those solutions be excluded by some physical restrictions on their behavior on the horizon or at infinity? Another question we would like to answer is whether those solutions correspond to linearization of some particular solution of the full Einstein equation in a similar way the conserved charges do?
}

An explicit reverse transformation for recovering the metric perturbation from the gauge-invariant functions was also provided.
As a side result --- we believe that the quasi-local gauge conditions, presented in section \ref{quasi-local}, could prove useful for analysis of problems of fall-off conditions, such as those encountered in \cite{rostworowski}.
The fall-off of metric and momentum components in this gauge is strictly dictated by the fall-off of the invariants, and as such {\alteration represents the truly physical behavior of the perturbation.}

A possible next step would be the inclusion of matter fields coupled to the perturbation.
{\alteration The fact that we do not use potential functions allows us to avoid some possible complications signaled in \cite{COS}. In fact, preliminary calculations suggest that adding a ``weak field'' to our model should be straightforward --- leading simply to appearance of some source terms in equations of motion. The introduction of a ``background'' energy-momentum tensor, however, could heavily disrupt our model, both by changing the background metric and affecting the gauge transformation of matter source terms in equations of motion. We plan to investigate this in our future works.}

\vspace{0.5 cm} \ \\
    {\noindent \sc Acknowledgements} This work was supported in part by
Narodowe Centrum Nauki (Poland) under Grant No. 2016/21/B/ST1/00940. One of the authors (P.W.) was
also supported by a special internal Grant for young researchers, provided by Center for
Theoretical Physics, PAS.
%}
%\pagebreak
\appendix
\section{Splitting equations with respect to the geometry of the two-spheres}
\label{splitting}
We present formulae for the gauge action, equations of motion and Gauss--Codazzi constraints, expressed in terms of quantities introduced in \ref{DataSplitting}.

\subsection{(2+1) splitting of the gauge}
\label{gaugesplitting}
Transformations of the metric (\ref{hxi}) split into radial, mixed and spherical part as follows:
\begin{align}
 \label{hxi33}
 h_{33} \quad \rightarrow &\quad h_{33} + \frac2{N}(N\xi_{3})_{,3}
\\ \label{hxi3A}
 h_{3A} \quad \rightarrow &\quad h_{3A} + \xi_{3,A}+ \xi_{A,3} -\frac2r \xi_A
 \\ \label{hxiAB}
 h_{AB} \quad \rightarrow &\quad h_{AB} + \xi_{A|| B}+ \xi_{B|| A} +\frac2r
\eta_{AB} \xi^3
\end{align}
Here ,,$||$'' denotes the two-dimensional covariant derivative with respect to $\eta_{AB}$.
The splitting of the transformation (\ref{pxi}) takes the form:
\begin{align}
\Pi ^{-1} P^{3}{_3} \quad \rightarrow &\quad \Pi ^{-1} P^3{_3}
 -\xi^{0||A}{_A} -\frac2r \xi^{0,3}
 \label{pxi33}  \\
 \label{pxi3A}
\Pi ^{-1} P_{3A} \quad \rightarrow &\quad \Pi ^{-1} P_{3A} +
\left[\frac{1}{N}(N\xi^0)_{,3}-\frac1r \xi^{0} \right]_{||A}
  \\
  \label{sxiAB}
\Pi ^{-1} S_{AB} \quad \rightarrow &\quad \Pi ^{-1} S_{AB}
  + \xi^0_{||AB} -\frac12\eta_{AB}\xi^{0||C}{_C}  \\
  \label{sxi}
  \Pi ^{-1} S \quad \rightarrow &\quad \Pi ^{-1} S
 -\frac2N(N\xi^{0,3})_{,3} -\frac2r \xi^{0,3}   -\xi^{0||C}{_C}
 \end{align}

Obtained expressions are formally similar to those for the Schwarzschild background \cite{praca}, just as it was in the case of constraint equations.

\subsection{(2+1) splitting of the equations of motion and constraints}
\label{EOMsplitting}

The equation of motion for the metric (\ref{hdot}) splits into following parts:
\begin{align}
\dot{h}_{33} \quad = &\quad \Pi^{-1}(P^3{_3}-S) + \frac 2{N}
({N} h_{03}),{_3} \label{h33}
\\
\dot{h}_{3A} \quad = &\quad 2f\Pi^{-1}P_{3A} + h_{03}{_{||A}} + h_{0A},{_3}
-\frac 2r h_{0A}   \label{h3A}
\\
\dot{h}_{AB} \quad = &\quad 2f\Pi^{-1}S_{AB}- \eta_{AB}\Pi^{-1}P^{33}     \notag
\\
&\quad +h_{0A||B} + h_{0B||A} + 2fr^{-1} \eta_{AB} h_{03}     \label{hAB}
\end{align}
Equation of motion for the ADM momentum is a bit more complicated:
\begin{align}
2\Pi^{-1}\dot{P}_{33}\quad = &\quad -\frac 1f h^0{_0}{^{||A}}_A -2r^{-1}h^0{_0},{_3} +
\frac 1f h_3{}^{3 ||A}{_A} + 2r^{-2}h^3{_3}                      \notag      \\
&\quad +  (H,{_3}- 2h_3{^{A}}{_{||A}} -2r^{-1}h_3{^3}),{_3}       \label{P33}     \\
&\quad +2r^{-1}(H,{_3}-2h_3{^A}{_{||A}}-2r^{-1}h_3{^3})        \notag
\\
2\Pi^{-1}\dot{P}_{3C}\quad = &\quad \left[\frac 1{\sqrt f} ({\sqrt f} h^0{_0}),{_3}
-r^{-1}h^0{_0}\right]_{ ||C}    \notag   \\
&\quad -\frac {f'}{2f} h^3{}_{3||C} +\frac{1}{2} (H,{_3}- 2h_3{^A}{_{||A}} -2r^{-1}h^3{_3})_{ ||C} \label{P3C}  \\
&\quad +h_{3C||A}{^{||A}} +\frac 1{r^2} h_{3C} - \chi^A{_{C||A}},{_3}  \notag
\\
2\Pi^{-1}\dot{S}_{AB} \quad = &\quad h^0{_{0||AB}} -\frac12 \eta_{AB}h^0{_0}^{||C}{_C}
+ h^3{_{3||AB}} -\frac12 \eta_{AB}h^3{_3}^{||C}{_C} \notag   \\
&\quad - (h^3{_{A||B}}+h^3{_{B||A}}-
\eta_{AB}h^{3C}{_{||C}}),{_3}+(f\chi^C{_B},{_3}\eta_{CA}),{_3} \notag \\
&\quad + \chi_{AB}{^{||C}}_{||C} - \chi^C{_{A||BC}} - \chi^C{_{B||AC}}  \label{SAB} \\
&\quad + \eta_{AB}\chi^{CD}{_{||CD}} +
\left(\frac 2{r^2}-2\Lambda\right)\chi_{AB}  \notag
\\
2\Pi^{-1} \dot S \quad = &\quad -2\sqrt{f} (\sqrt{f} h^0_{0,3})_{,3}-  h^0_{0,3} \left(\frac 2r - 2r\Lambda\right)
-h_0^{0||A}{_A} + (h^3{_3} +H)^{||A}{_A}  \notag \\
&\quad +\frac2{r^2}(h^3{_3} +H) -\left(\frac{12m}{r^3}+2\Lambda\right) h^3{_3} \notag   \\
&\quad +\frac{2m}{r^2}h^3{_{3,3}} +  f(H,{_3}- 2h_3{^{A}}{_{||A}} -2r^{-1}h_3{^3}),{_3}  \label{Sdot} \\
&\quad+4\frac{f}r (H,{_3}-2h_3{^A}{_{||A}}-2r^{-1}h_3{^3}) -2\chi^{CD}{_{||CD}}
  \notag
\end{align}
Vector constraint (\ref{wwl}) splits into:
 \begin{equation}
\frac 1{\sqrt f}({\sqrt f}P^3{_3}),{_3} +  P_3{^A}{_{||A}} - r^{-1}S = 0
\label{ww3}
\end{equation}
\begin{equation}
P^3{_A},{_3} + S_A{^B}{_{||B}}+\frac{1}{2} S_{||A} = 0
\label{wwA}
\end{equation}
And the scalar constraint (\ref{wsl}) gives:
\begin{equation}
\begin{aligned}
 h ^{|l}{_l} - h^{kl}{_{|kl}} +h^{kl} \Rthree_{kl}
 = \frac{\sqrt f}{r^3}\left[ {r^2}{\sqrt f}
 (rH,{_3}-2rh_{3A}{^{||A}}-2h_3{^3})\right],{_3}  \\
 +h^3{_{3 ||A}}{^A}+2r^{-2}h^3{_3} -\frac{6m}{r^3}h^3{_3}
 +\frac{1}{2}H^{||C}{_C}+r^{-2}H
  - \chi^{AB}{_{||AB}} = 0   \label{wsk}
\end{aligned}
\end{equation}

 \section{Derivatives of a diagonal metric}
\label{metryka}
Let us consider a metric in the following form:
\begin{displaymath}
\eta_{\mu\nu}=-f(r)\dd[t]^2 + \frac{1}{f(r)} \dd[r]^2+ r^2\left[\dd[\vartheta]^2+\sin^2\vartheta\dd[\varphi]^2 \right]
\end{displaymath}
The non-zero Christoffel symbols for the given metric are:
\begin{alignat*}{3}
\kri{t}{tr}= &\quad \frac{f'}{2f}&\qquad \kri{r}{\vartheta\vartheta} = & -f r &\qquad  \kri{\vartheta}{\varphi\varphi}=&-\sin\vartheta\cos\vartheta  \\
\kri{r}{tt} = &\quad \frac{1}{2}f f' &\qquad \kri{r}{\varphi\varphi}=& - f r \sin^2\vartheta &\qquad \kri{\varphi}{r\varphi}=&\quad \frac{1}{r} \\
\kri{r}{rr} = & -\frac{f'}{2f}&\qquad  \kri{\vartheta}{r\vartheta}=&\quad \frac{1}{r}  &\qquad  \kri{\varphi}{\vartheta\varphi}=&\quad \ctg \vartheta
\end{alignat*}
By setting $f(r)\equiv 1$ we obtain Christoffel symbols for a flat Minkowski spacetime.

Taking $f(r)\equiv 1-\frac{2m}{r}$ gives us their values for the Schwarzschild metric, in standard coordinates.

Setting $f(r)=1-\frac{2m}{r}-\frac{r^2}{3}\Lambda$ produces Christoffel symbols for the Kottler metric.

Let us note that the surface defined by $t=\,$const. always possesses a vanishing external curvature,  for any form of the function $f(r)$, as all Christoffel symbols with indices $\kri{t}{kl}$ are equal to zero.

In this paper we use the following convention for the Riemann tensor:
\begin{equation*}
\Rie{\mu}{\nu\alpha\beta}v^{\nu}=(\nabla_{\alpha}\nabla_{\beta}-\nabla_{\beta}\nabla_{\alpha})v^{\mu} = 2v^\mu{}_{;[\beta\alpha]}
\end{equation*}

Its components for the given metric are:
\begin{alignat*}{3}
\Rie{t}{rtr}=&-\frac{f''}{2f}&\qquad \Rie{t}{\vartheta t \vartheta}=&-\frac{f'r}{2} &\qquad \Rie{t}{\varphi t \varphi}=&-\frac{f'r}{2}\sin^2 \vartheta\\
\Rie{r}{\vartheta r \vartheta}=& -\frac{f'r}{2}&\qquad \Rie{r}{\varphi r \varphi}=& -\frac{f'r}{2}\sin^2 \vartheta &\qquad \Rie{\vartheta}{\varphi \vartheta \varphi} =&(1-f)\sin^2 \vartheta
\end{alignat*}

It is also useful to know the values of Christoffel symbols and the Riemann tensor in lower dimension, for the restriction of $\eta_{\mu\nu}$ to the hypersurface $t=\,$const. Because our metric is diagonal, the three-dimensional Christoffel symbols will simply be equal to their four-dimensional counterparts with analogous indices. For this particular form of the metric (with $f$ depending solely on $r$), a similar relation exists between the values of the three- and four-dimensional Riemann tensor.

For the readers convenience, we provide explicit formulae for the Ricci tensor and scalar in three and four dimensions.
\begin{alignat*}{2}
\Rfour^t{}_t \,&=\Rfour^r{}_r=-\frac{f''}{2}-\frac{f'}{r} & \qquad \Rfour^\vartheta{}_\vartheta  & = \Rfour^\varphi{}_\varphi = \frac{1-f}{r^2}-\frac{f'}{r}\\
\Rfour \;\;\, & = 2\frac{1-f}{r^2} - 4\frac{f'}{r} - f'' \\
\Rthree^r{}_r&=-\frac{f'}{r} & \quad \Rthree^\vartheta{}_\vartheta  & = \Rthree^\varphi{}_\varphi = \frac{1-f}{r^2} - \frac{f'}{2r}\\
\Rthree \;\;\, & = 2\frac{1-f}{r^2}-2\frac{f'}{r}
\end{alignat*}
Taking $\displaystyle f(r)=1-\frac{2m}{r}-\frac{r^2\Lambda}{3}$, we obtain:
\begin{alignat*}{2}
\Rfour_{\mu\nu} & = \Lambda \eta_{\mu\nu} & \qquad \Rfour \quad &= 4 \Lambda \\
\Rthree^r{}_r&= \frac{2}{3}\Lambda - \frac{2m}{r^3} & \qquad \Rthree^\vartheta{}_\vartheta & = \Rthree^\varphi{}_\varphi = \frac{2}{3} \Lambda + \frac{m}{r^3} \quad \Rthree = 2\Lambda
\end{alignat*}

%\newpage

\end{document}